\documentclass[onecolumn]{aa}
\usepackage{natbib}
\usepackage{graphicx}

\bibpunct{(}{)}{;}{a}{}{,}

\newcommand\bb[1] {   \mbox{\boldmath{$#1$}}  }
\newcommand\del{\bb{\nabla}}
\newcommand\bcdot{\bb{\cdot}}
\newcommand\btimes{\bb{\times}}

\begin{document}


\title{Properties and stability of freely propagating nonlinear
  density waves in accretion disks.}
\author{S\'ebastien Fromang  and John Papaloizou}

\offprints{S.Fromang}

\institute{Department of Applied Mathematics
and Theoretical Physics, University of Cambridge, Centre for
Mathematical Sciences, Wilberforce Road, Cambridge, CB3 0WA, UK \\ \email{S.Fromang@damtp.cam.ac.uk}}

\date{Accepted; Received; in original form;}

\label{firstpage}

\abstract
{}
{In this paper, we study the propagation and stability of nonlinear
  sound waves in accretion disks.}
{Using the shearing box approximation, we derive the form of these waves
   using a semi-analytic approach and  go on to  study their stability. The results are 
  compared   to those of numerical simulations   performed     using finite difference
 approaches  such as employed by ZEUS  as well as 
  Godunov methods.}
{When the wave frequency is between $\Omega$ and $2 \Omega$ (where
  $\Omega$ is the disk orbital angular velocity), it   can couple resonantly with a
    pair of linear inertial waves and thus  undergo  a parametric
  instability. Neglecting the disk vertical stratification, we derive
  an expression for  the growth rate when the amplitude of
  the background wave is small. Good agreement is found with the
  results of numerical simulations performed both with finite
  difference  and Godunov codes. During the nonlinear phase
  of the instability, the flow remains well organised if the amplitude
  of the background wave is small.  However, strongly nonlinear waves
  break down into turbulence. In both
  cases, the  background wave is damped and the disk 
  eventually   returns to a  stationary state. Finally, we demonstrate that
  the instability  also
   develops when density stratification is taken into account and so is robust.}
{This destabilisation of freely propagating nonlinear sound waves may
  be important   for understanding the complicated behaviour of density waves
in  disks that are unstable through the effects of self-gravity or magnetic fields
 and is likely to affect the
propagation of waves that are tidally excited by objects such as a protoplanet or companion  perturbing a
protoplanetary disk.
The nonlinear wave solutions described here as well as their stability
properties were also   found to be useful for testing and comparing the performance
of different numerical codes.}
\keywords{Accretion, accretion disks - Methods: numerical}

\authorrunning{S.Fromang \&  J.Papaloizou}
\titlerunning{Nonlinear sound waves propagation}
\maketitle

\section{Introduction} \label{Intro}

Waves and oscillations are  expected to be found in accretion disks as they can be
excited by a very large variety of phenomena. For example, tidal excitation can be
caused by the influence of a binary
companion  \citep{artymowicz&lubow94} or
by a planet orbiting inside the disk
\citep{gt79,gt80,nelsonetal00}. Spiral density waves can be excited as
a result of gravitational instability   when  
 the disk mass is large enough \citep{lin&shu64,toomre81,shu92}. It has also been suggested
that waves are responsible for the puzzling  phenomenon of QPOs
\citep{kato01,arrasetal06}. Recent numerical simulations of MHD
turbulence resulting from the magnetorotational instability
\citep[MRI;][]{balbus&hawley98} have also indicated that density waves could be
excited by the turbulent fluctuations of the flow
\citep{papaloizouetal04,gardiner&stone05b}.

  There have been many studies of  the properties of the large
variety of waves that can propagate in disks. Most of the early work
was done using simplifying assumptions and restricting the analysis to the
linear case
\citep{linetal90a,linetal90b,lubow&pringle93,korycanskyetal95,lubow&ogilvie98,ogilvie&lubow99}.
Some other work extended these results to the nonlinear regime, either
 semi-analytically, but 
retaining  terms up to  second order in  an expansion in  powers of the amplitude \citep{larson90},
or numerically in order to study wave dissipation
\citep{bateetal02,torkelssonetal00}. 

Despite all these efforts, a complete semi-analytic solution to the wave
equation, valid in the nonlinear regime for large amplitude, is still lacking. This would
not only be useful   for studying   the properties of waves, but would also provide a
valuable test for numerical codes that are more and more extensively used in
the accretion disk community. The goal of this paper is to derive
such a solution and to study its properties and stability. Of course,
the need to derive a solution  that can be found using semi-analytic methods imposes strong
constraints: we use an isothermal equation of state, adopt the
shearing box approximation \citep{goldreich&lyndenbell65} and restrict the analysis
to the case where there is no variation in the direction of rotation or axisymmetry.
 But all these approximations simplify the analysis which
can then be done including nonlinearities.

In  this paper we are concerned with parametric
instability occurring for non linear density waves treated in a local shearing
box approximation. However,
parametric instability appears also to be generic  for global  non axisymmetric,
non magnetic disk flows such as may occur   when the disk is 
either eccentric
\citep{goodman93,ogilvie01,pap05a,pap05b} or warped \citep{torkelssonetal00}.
It is also related to the elliptical instability observed in rotating flows
that is apparently connected to turbulent breakdown \citep{kerswell02}.
Such turbulent breakdown does seem to occur for some of the large amplitude non circular
flows studied here.
All of this makes the phenomenon an important one for study in the general
context of the dynamics of accretion disks.
Being a local instability that occurs through the interaction of a dense
spectrum of inertial waves with a primary flow,  modes with arbitrarily small
length scales, depending ultimately on the dissipative
properties of the system  may be excited, providing a challenge for numerical simulations.

The plan of the paper is as follows. In section~\ref{analytic}, we
derive the equations describing the propagation of nonlinear density
waves in the shearing box approximation. The solutions to these
equations are compared with the results of 1D numerical solutions. In
section~\ref{stability_section}, we study their stability. When the
frequency of the wave lies in the range $[\Omega,2\Omega]$, where
$\Omega$  is the orbital angular velocity, it can interact resonantly with two linear
inertial waves 
(whose individual frequencies, being bounded by $\Omega,$  combine to equal the wave frequency) and be
unstable to a parametric instability
\citep{goodman93,torkelssonetal00,pap05b}. Neglecting
vertical density stratification in this section, we derive an
expression for the growth rate of the instability and describe the
properties of the eigenmodes. These results are
compared in section~\ref{insta_num} with 2D numerical
simulations of vertically unstratified shearing boxes. Both finite
difference  schemes as used in ZEUS and NIRVANA (see below)  as well as  Godunov methods are 
shown to give results that  agree very well with each other and with the results of the
linear analysis. Finally, in section~\ref{strat_section}, we
investigate the effect of vertical stratification, showing the occurrence
of parametric instability in that case and relating the results to the
unstratified case.   Finally we   discuss our results
and summarize conclusions in 
section~\ref{discussion_section}.

\section{Free waves}
\label{analytic}

\subsection{Basic equations }
In common with many studies of the local dynamics of accretion disks orbiting about a central mass,   we adopt the shearing box approximation
\citep{goldreich&lyndenbell65}. In this approximation, a reference frame
 rotating with a fixed angular
velocity $\Omega,$ which corresponds to the Keplerian angular velocity at a specified radius $R_0,$
is adopted. The origin, at the centre of the box, is
 located at some specified point on the circle of radius $R_0$ and a system of
 Cartesian coordinates $(x,y,z)$
with associated unit vectors $(\bb{i},\bb{j},\bb{k})$ is used.
Gravity is due to the central object only  and an expansion of the gravitational acceleration
correct to first order in $x$ and $z$ is adopted. 
 Then the equations of continuity and momentum conservation can be written in the form 
\begin{eqnarray}
\frac{\partial \rho}{\partial t} + \del \bcdot (\rho \bb{v}) &=& 0 \, ,\\
\frac{\partial \bb{v}}{\partial t} + (\bb{v} \bcdot \del) \bb{v} + 2
\bb{\Omega} \btimes \bb{v} &=& -\frac{1}{\rho} \del P + 3 \Omega^2 x
\bb{i} - \Omega^2 z \bb{k} \, ,\label{BEQ}
\end{eqnarray}
where  $\rho$ is the density,
$\bb{v}$ the velocity and $P$ is the
pressure which is  related to the density by an equation of state (EOS). In
this paper we restrict consideration to an isothermal EOS,
for which
\begin{equation}
P=\rho c^2 \, ,
\end{equation}
where $c = \Omega H$ is the isothermal sound speed with $H$ being
the scale height. In the absence of waves, the
equilibrium solution  takes the form
\begin{eqnarray}
\rho &=& \rho_0 \exp\left(-\frac{z^2}{2H^2}\right)\, ,\\
\bb{v} &=& -\frac{3}{2} \Omega x \bb{j}\, ,
\end{eqnarray}
where $\rho_0,$ being the midplane density, is a constant taken to be
unity in the following. In some cases we neglect the vertical
component of the gravitational acceleration, $-\Omega^2 z {\bf {\hat
    k}},$ in Eq.~(\ref{BEQ}). Then vertical stratification is also
neglected and $\rho = \rho_0$ throughout.

\subsection{The wave equation}
\label{wave_equation}

Initially, for simplicity, we neglect vertical  stratification and assume that the state variables
do  not vary along the direction of rotation or the vertical direction 
 and thus depend only on  $x$ and $t.$
To derive the equations that describe wave propagation in this system, we adopt
a Lagrangian description. We define  $X(X_0,t)$ to be  the position, or $x$ coordinate,
 of a fluid element that would be at the fixed location $x=X_0$ in the absence of any
disturbance. The $y$ component of the momentum
conservation equation can be written 
\begin{equation}
\frac{D v_y}{D t} + 2 \Omega \frac{D X}{D t} = 0 \, ,
\label{y equation}
\end{equation}
where the convective derivative is
\begin{equation}
\frac{D}{Dt}=\frac{\partial }{\partial t} + v_x \frac{\partial
}{\partial x} \, ,
\label{conv_der}
\end{equation}
and we have used the fact that
\begin{equation}
v_x=\frac{D X}{Dt} \, .
\label{vx_def}
\end{equation}
Eq.~(\ref{y equation}) can be integrated  to give
\begin{equation}
v_y+2\Omega X =  f(X_0) \, .
\label{vy relation}
\end{equation}
where $f(X_0)$ is a function of $X_0$ to be determined.

Applying the same procedure to the equation of motion in the $x$ direction, we find
\begin{equation}
\frac{D v_x}{D t} + \Omega^2 X = -\frac{1}{\rho} \frac{\partial
  P}{\partial x} + 2 \Omega f(X_0) \, ,
\label{vx inter}
\end{equation}
where we have  used Eq.~(\ref{vy relation}) to eliminate $v_y$. 
In the
absence of disturbance, $v_x=0$ and the pressure is uniform. 
This means we must have 
\begin{equation}
f(X_0)=\frac{1}{2} \Omega X_0 \, .
\end{equation}
Thus Eq.~(\ref{vx inter}) becomes
\begin{equation}
\frac{D^2 X}{D t^2} + \Omega^2 (X-X_0) = -\frac{c^2}{\rho}
\frac{\partial \rho}{\partial x} \, .
\label{x inter}
\end{equation}
In order to express the density in terms of $X,$ we use the
conservation of mass. This gives  
\begin{equation}
\rho \frac{\partial X}{\partial X_0} \equiv \rho\left( 1+{\partial \xi \over \partial X_0}\right)= \rho_0  \, ,
\label{mass cont}
\end{equation}
where we have  introduced  the displacement $\xi(X_0,t)$ through the
relation $X(X_0,t)=X_0+\xi(X_0,t)$.

\subsubsection{Travelling waves}
Since the natural boundary conditions for the shearing box are
periodic in shearing coordinates, we consider travelling wave  solutions
that depend only on the phase $\Phi=X_0-Ut$, where $U$ is a constant
the phase velocity. Then we have
$\xi=\xi(\Phi)$. Using Eq.~(\ref{mass cont}) to eliminate the
density, we obtain 
\begin{equation}
U^2 \frac{d^2 \xi}{d\Phi^2} + \Omega^2 \xi = \frac{c^2}{ ( 1+
  \frac{d\xi}{d\Phi}  )^2} \frac{d^2 \xi}{d\Phi^2} \, .
\label{wave equation}
\end{equation}
which is the governing equation that describes travelling wave propagation. It has been
derived assuming no magnetic field is present. We comment here that a
 vertical or an azimuthal field (or both) that depend only on $\Phi$ can
easily be incorporated into the above analysis. However, these details
are beyond the scope of this paper.
We comment also that there are some similarities of the above analysis
to that of \citet{larson90}. He considered travelling waves
in a disc though not explicitly in a shearing box. However, his analysis
focused on spiral waves and  only terms of second order
or lower in the wave amplitude were retained.  This  is inexact
and inapplicable to the large amplitude waves considered here.

\subsection{First integral of motion}

\begin{figure}
\begin{center}
\includegraphics[scale=0.5]{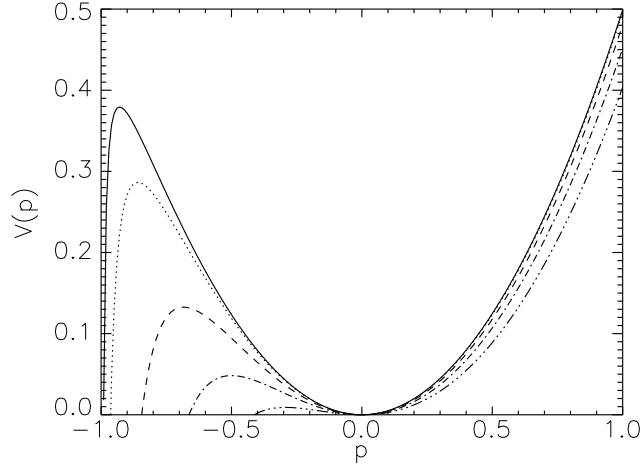}
\caption{Profile of the function $\tilde{V}(p)$ for different values
  of $(c/U)^2$. The solid, dotted, dashed, dotted-dashed and
  dotted-dotted-dotted-dashed curves respectively correspond to
  $(c/U)^2=0.005,0.02,0.1,0.25,0.5$.}
\label{potential}
\end{center}
\end{figure}

By considering a first integral of Eq.~(\ref{wave equation}), it
is possible to interpret the problem as corresponding to the motion
of a particle in a potential well. To see this, let $p$ be defined through
\begin{equation}
p=\frac{d\xi}{d\Phi} \, .
\label{p_def}
\end{equation}
Using this, Eq.~(\ref{wave equation}) can be written 
\begin{equation}
\left(U^2 -\frac{c^2}{(1+p)^2}\right) p\frac{dp}{d\xi}+\Omega^2
\xi =0 \, .
\end{equation}
which can be integrated with respect to $\xi$ to give
\begin{equation}
\frac{1}{2}\Omega^2\xi^2 =
\frac{1}{2\Omega^2}\left(U^2 -\frac{c^2}{(1+p)^2}\right)^2 \left(\frac{dp}{d\Phi}\right)^2 =U_0^2- U^2\tilde{V}(p) \, ,
\label{first int}
\end{equation}
where the dimensionless function $\tilde{V}(p)$ is given by
\begin{equation}
\tilde{V}(p)= \frac{p^2}{2}+(c/U)^2\left(\frac{p}{1+p}-\ln(1+p)\right)
\end{equation}
and $U_0$  is a constant.
Equation~(\ref{first int}) shows 
by analogy with classical mechanics that the evolution of $p$ as a function of $\Phi$ 
can be periodic,  consisting of  oscillations in the wells of the function
$\tilde{V}(p),$ should they exist.  The bounds of such oscillations
are determined by the condition $\tilde{V}(p) =
(U_0/U)^2$.  By definition, $\xi$ and $p$ should
oscillate around $p=0$ which should be a minimum of $V$ and be such that $p > -1$.
Such a situation is  
 only possible if $c / U  \le 1. $
 The form of $\tilde{V}(p)$ is shown in figure
\ref{potential}. The different curves correspond to $(c/U)^2=0.005$
({\it solid line}), $0.02$ ({\it dashed line}), $0.1$ ({\it dotted line}),
$0.25$ ({\it dotted-dashed line}) and $0.5$ ({\it dotted-dotted-dotted-dashed
  line}). The form of $\tilde{V}(p)$ is similar for all these values of $(c/U)^2.$
It increases for $-1 < p < -1+c/U$, decreases for $-1+c/U < p < 0$
 and increases again for $p>0$. 
The  well allows  $p$  to oscillates around
$0$. Figure~\ref{potential} also shows that the amplitude of the
oscillations of p is bounded such that $|p|<1-c/U$

\subsection{The linear limit}

When $|p|$ is small, the wave becomes linear. This happens when $|p| \ll
1-c/U$.  In this limit the potential is parabolic, being given by 
\begin{equation}
\tilde{V}(p) \sim\frac{p^2}{2} \left( 1-(c/U)^2 \right) \, .
\end{equation}
The system then behaves as an harmonic oscillator. 
In this linear limit, Eq.~(\ref{wave equation}) becomes
\begin{equation}
(U^2-c^2)\frac{d^2\xi}{d\Phi^2}+\Omega^2\xi=0 \, .
\end{equation}
The solution is
\begin{equation}
\xi=\xi_0 \cos (kX_0-\omega t) + \xi_1 \sin (kX_0-\omega t)  \, ,
\end{equation}
with
$k = \Omega / \sqrt{U^2-c^2},$
and $\omega = kU$ which together imply the standard dispersion relation
for linear sound waves modified by rotation in the form
$\omega^2=k^2 c^2 + \Omega^2.$

\subsection{The effect of vertical stratification}

The same equations as above apply when the disk is vertically
stratified. The nonlinear wave has exactly the same
properties in that case. Both  the perturbed and unperturbed  density
are  simply  multiplied by a factor $\exp \left(-\frac{z^2}{2H^2}\right)$
which contains the entire vertical dependence.

\subsection{Solutions of the wave equation}
\label{ode_sol}

\begin{figure}
\begin{center}
\includegraphics[scale=0.5]{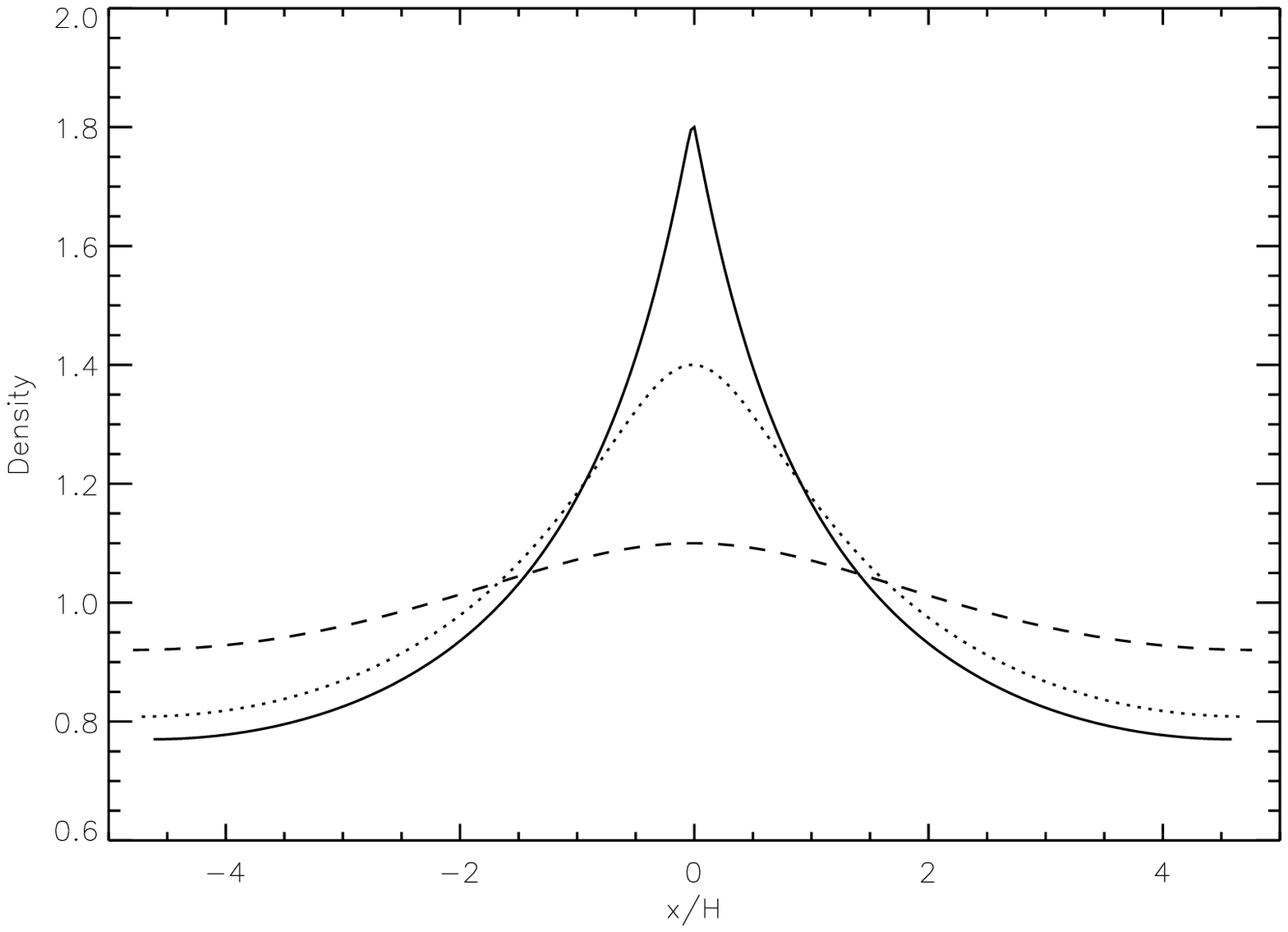}
\includegraphics[scale=0.5]{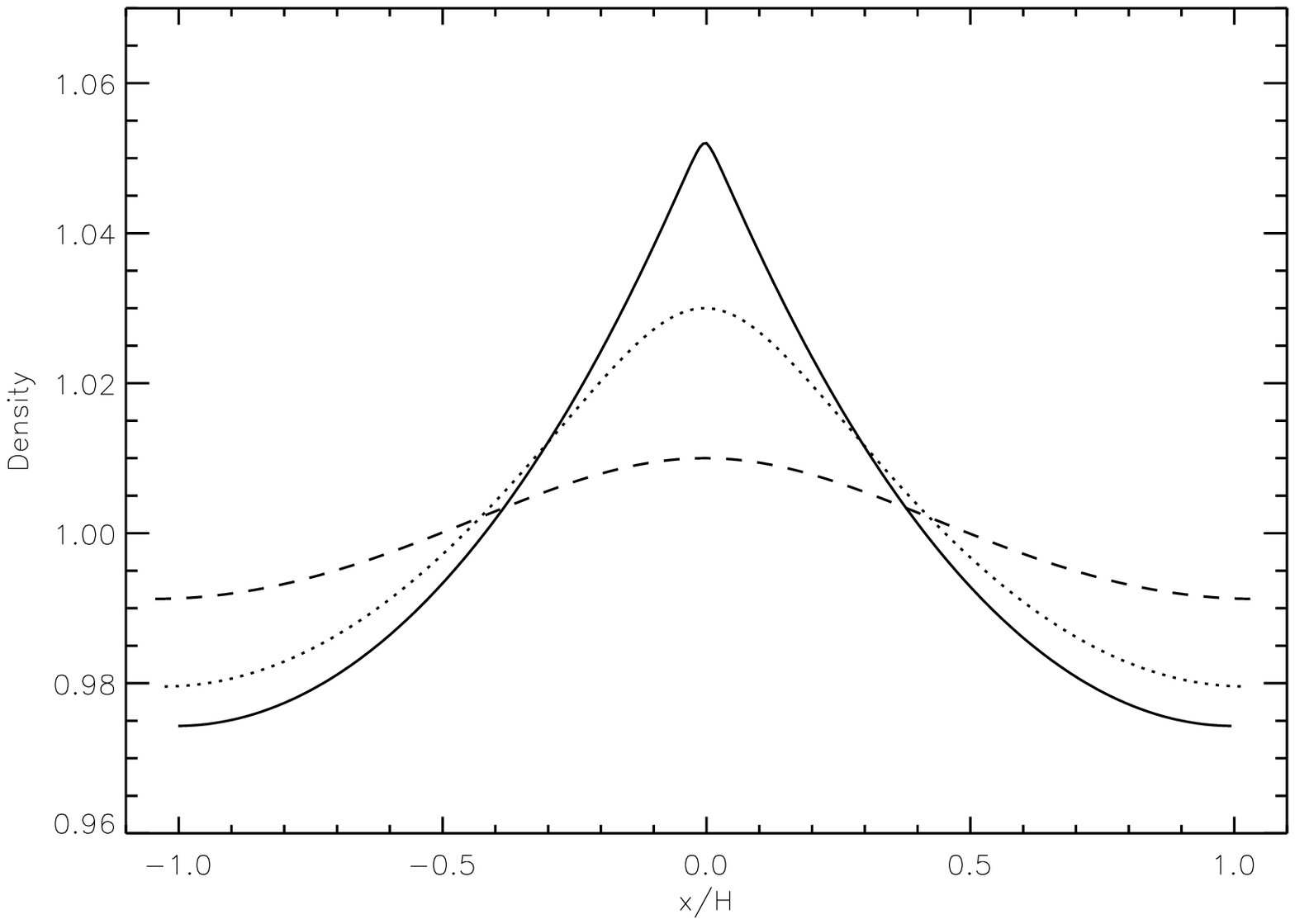}
\caption{Left panel: radial profile of the density for a
    wave of velocity $(c/U)^2=0.3$, for different wave amplitudes:
    $\rho_{max}/\rho_0 =1.1$ ({\it dashed line}), $1.4$ ({\it dotted line})
    and $1.8$ ({\it solid line}). The solution shows a cusp
    around its maximum as the amplitude of the wave is
    increased. Right panel: same as the left panel, but for
    $(c/U)^2=0.9$ and $\rho_{max}/\rho_0=1.01$ ({\it dashed line}), $1.03$
    ({\it dotted line}) and $1.052$. Note the smaller wave amplitude
    and the smaller periods in that case compared to the case $(c/U)^2=0.3$.}
\label{wave_struct}
\end{center}
\end{figure}

Using Eq.~(\ref{p_def}), the wave equation (\ref{wave equation}) can
be written as a system of two coupled first order differential equations:
\begin{eqnarray}
\frac{d\xi}{d\Phi} &=& p \, . \label{ode1} \\
\frac{dp}{d\Phi} &=& - \left( \frac{\Omega}{c} \right)^2 \frac{(c/U)^2
  \xi}{1-\frac{(c/U)^2}{ ( 1+p )^2}}\, . \label{ode2}
\end{eqnarray}
Once the phase $\Phi$ and displacement $\xi$ are normalized
by the natural shearing box length scale $\Omega/c,$  a solution of the wave 
equation  is specified by the parameter $c/U$ and the
maximum value of the density in units of $\rho_0,$  $\rho_{max}/\rho_0.$ The latter fixes the
initial conditions for the integration, which is performed until the
solution reaches the next maximum. This procedure gives the spatial
period $T_w$ of the wave, or, equivalently, its wavenumber $k_{x,w}=2\pi/T_w.$

The solutions we obtained are illustrated in figure
\ref{wave_struct}. The left panel shows the structure of the
wave when $(c/U)^2=0.3$ for different values of the relative wave amplitude
$\rho_{max}/\rho_0 = 1.1$, $1.4$ and $1.8$ (we note that the maximum possible free wave
amplitude  is $\rho_{max}/\rho_0 = U/c = 1.825$
for these parameters). For these three values, we respectively found
$T_w=9.58H$, $9.43H$ and $9.23H$. As the amplitude of the wave is
increased, nonlinearity becomes more and more important and a cusp
eventually forms around the location of the density maximum. The right
panel shows the structure of the wave when
$(c/U)^2=0.9$. Again, three different wave amplitudes illustrate the
transition from linear to nonlinear wave profiles. These are given by
$\rho_{max}/\rho_0=1.01$,$1.03$ and $1.052$. We note that similar waveforms
occur for both values of $c/U$.  However, for  corresponding waveforms, when
$(c/U)^2=0.9$, the amplitudes and periods of the wave are much smaller
than when $(c/U)^2=0.3.$
 
\subsection{Numerical tests}
\label{tests}

\begin{figure}
\begin{center}
\includegraphics[scale=0.33]{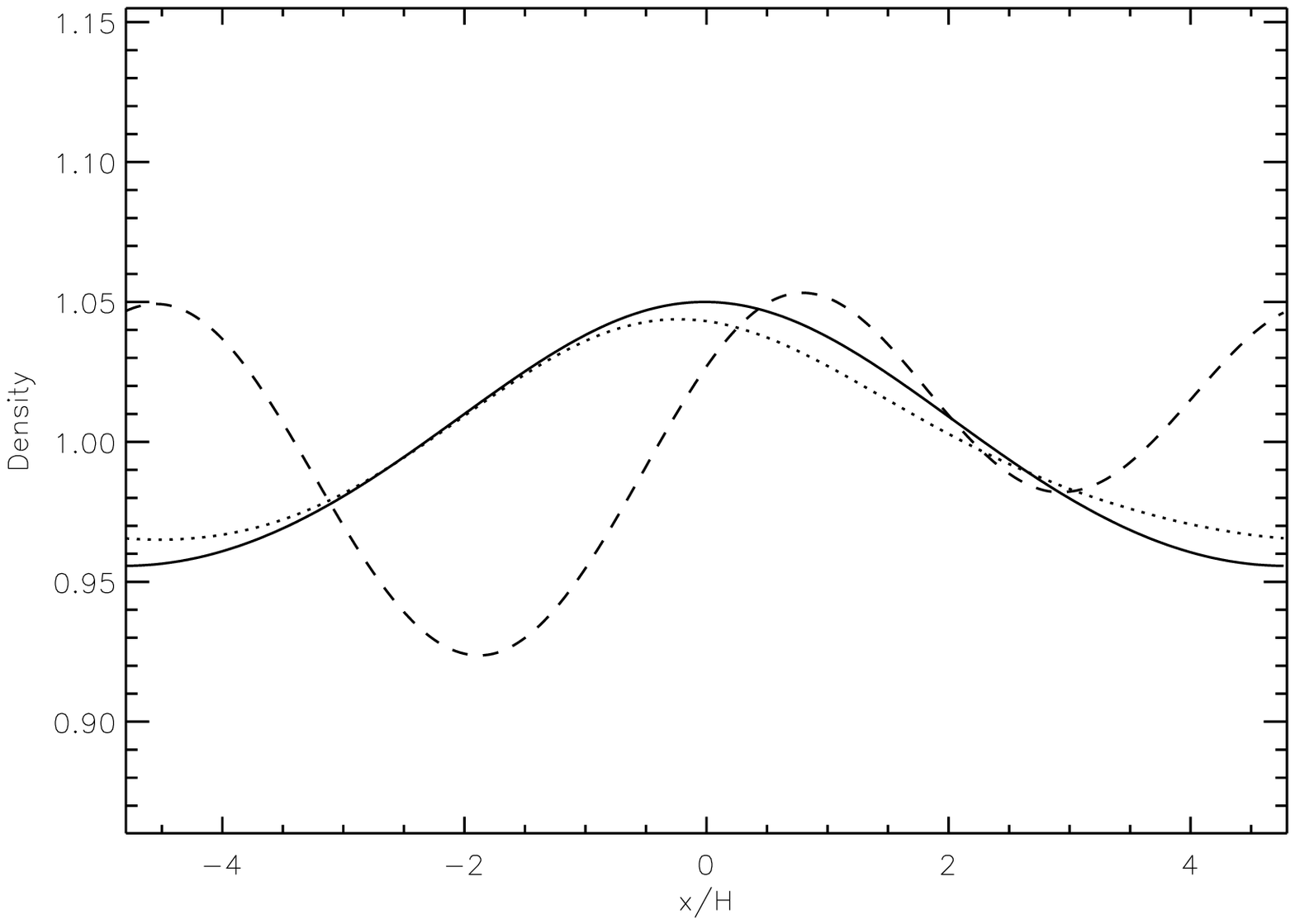}
\includegraphics[scale=0.33]{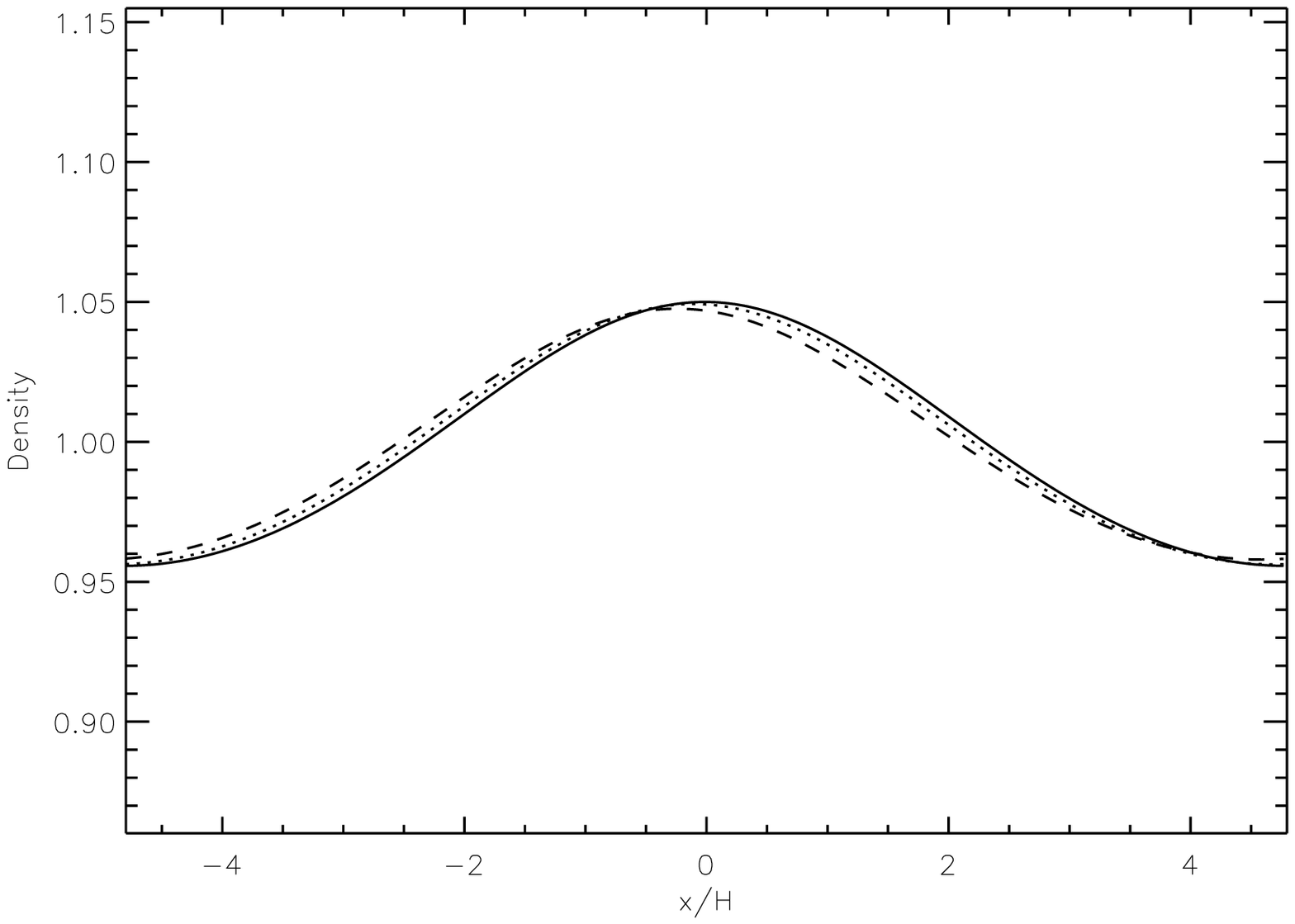}
\includegraphics[scale=0.33]{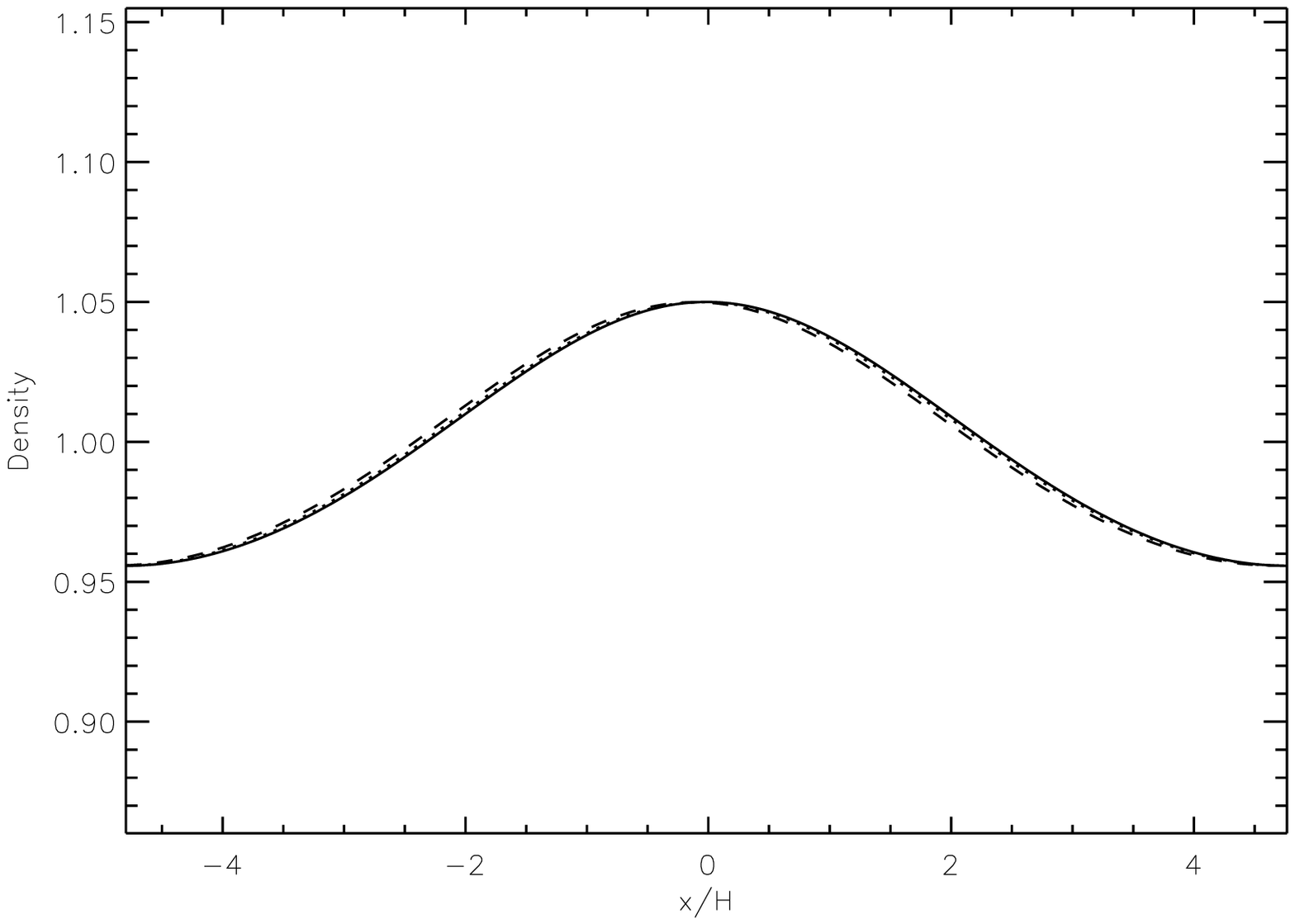}
\includegraphics[scale=0.33]{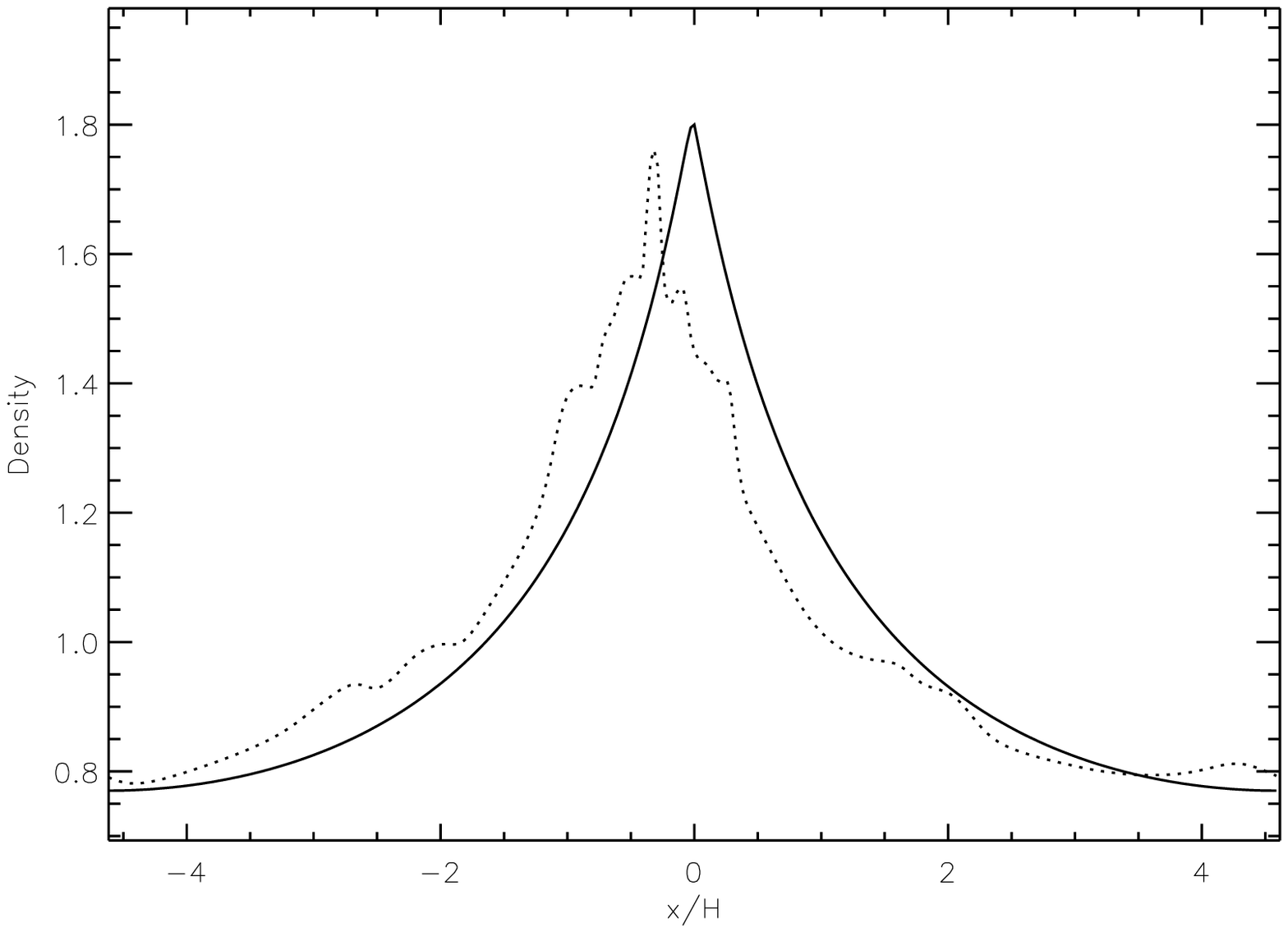}
\includegraphics[scale=0.33]{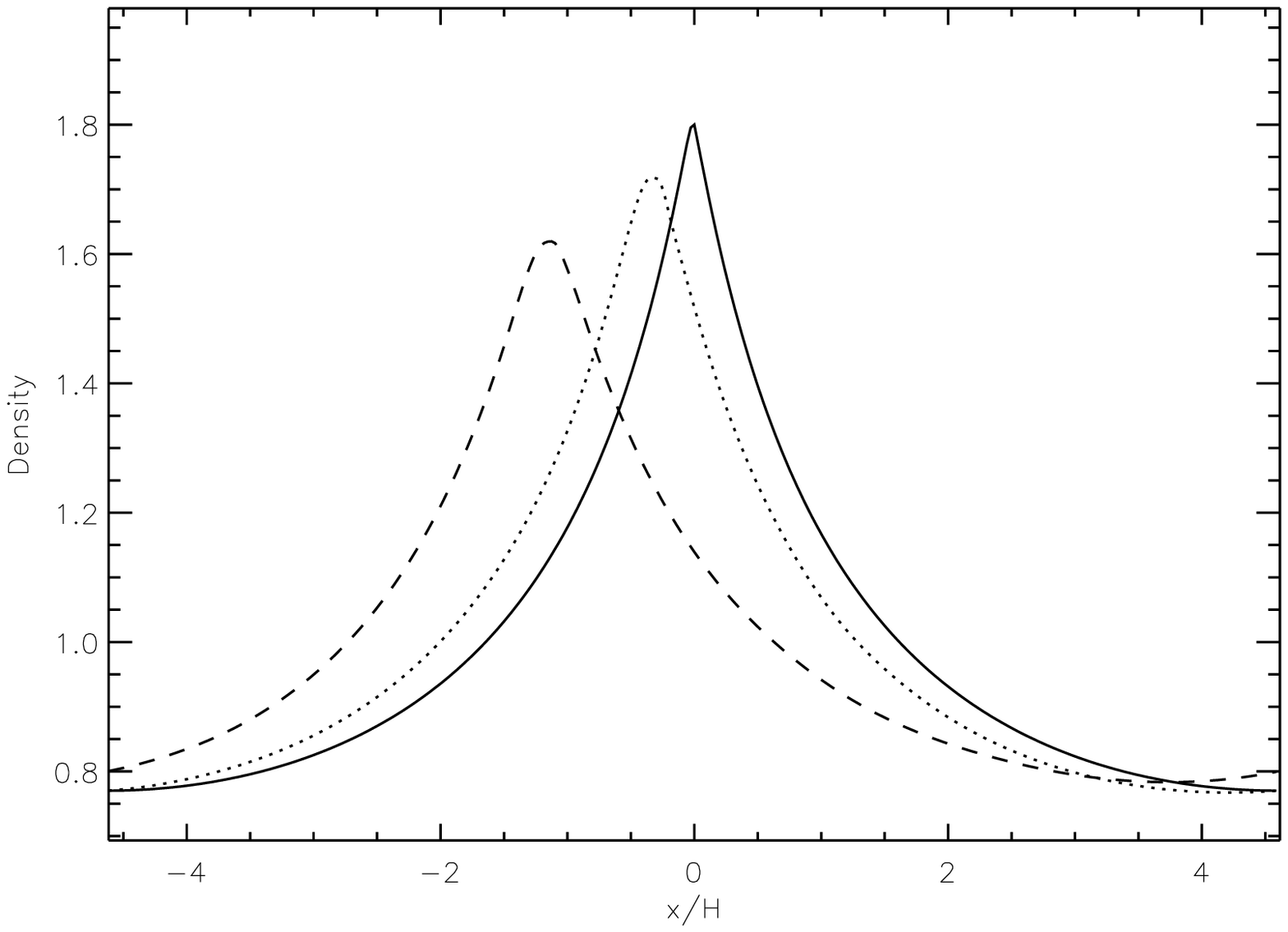}
\includegraphics[scale=0.33]{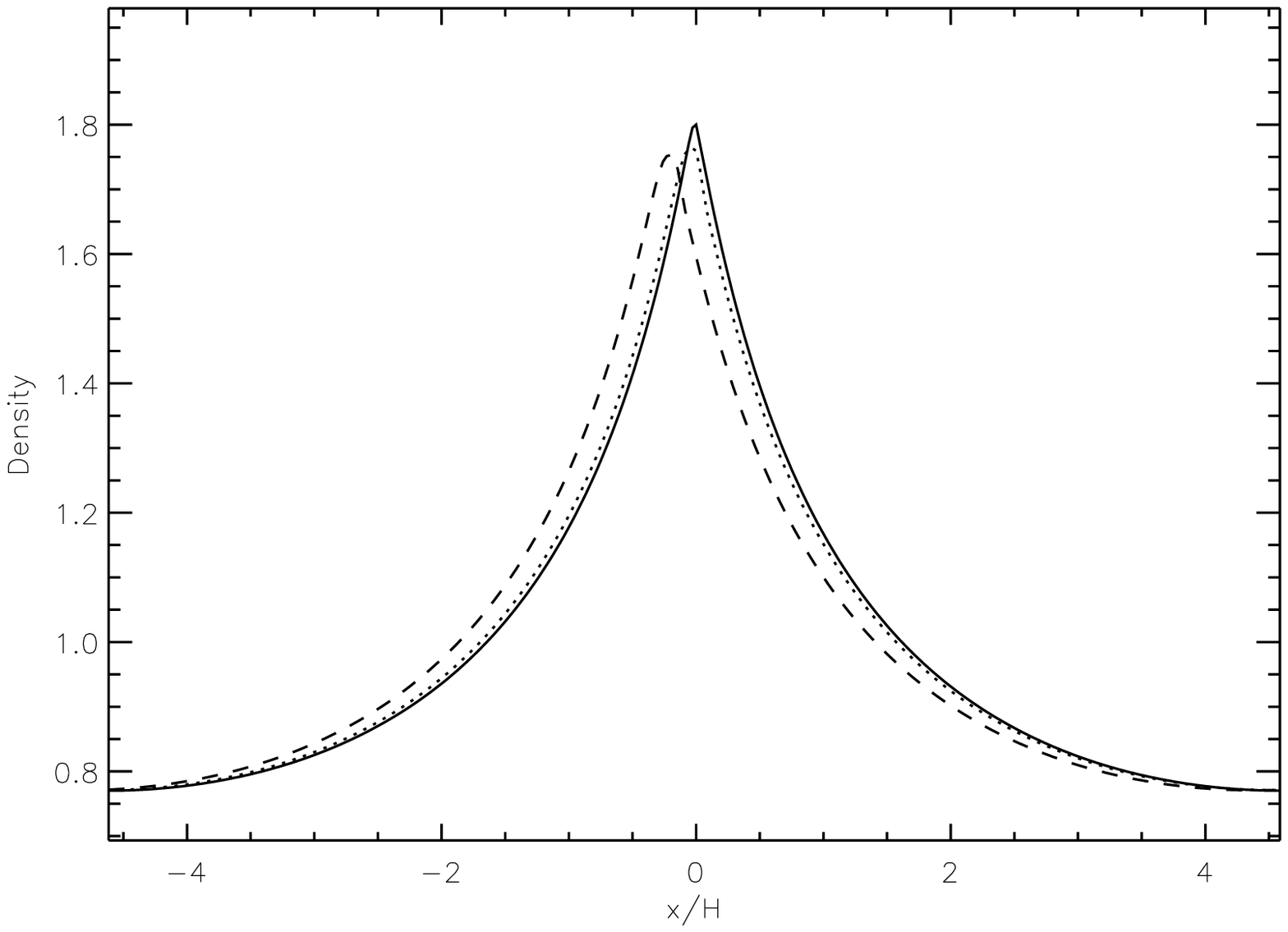}
\caption{Results of the numerical tests illustrating the propagation
  of 1D nonlinear waves in the shearing box. The parameters are such
  that $(c/U)^2=0.3$ and $\rho_{max}/\rho_0=1.05$ ({\it upper row}) and
  $\rho_{max}/\rho_0=1.8$ ({\it lower row}). The resolution is with $320$ grid
  points in the radial direction. Upper and lower left 
  panels show results obtained with ZEUS using a Courant number
  $C_0=0.5$, the upper and lower central panels are obtained with ZEUS using $C_0=0.1$
  The  upper and lower right  panels are obtained with RAMSES using
  $C_0=0.7$. For the upper panels, the plots are for times
  (measured in  cycles or wave periods, with one cycle or period being
  the time for the wave to cross the box)
  $t=0$ ({\it solid line}), $t=250$ ({\it dotted line}) and $t=500$
  ({\it dashed line}). For the lower panels, the plots are for 
  times $t=0$ ({\it solid line}), $t=30$ ({\it dotted line}) and
  $t=60$ ({\it dashed line}). Note that the profiles at $t=0$ are
  computed by solving the wave equation as described in section~\ref{wave_equation}.}
\label{1d_tests}
\end{center}
\end{figure}

The results presented in section \ref{analytic} can be used to provide   1D tests
for hydrodynamic codes. Here we present such test calculations
performed with two finite difference
codes, ZEUS \citep{hawley&stone95} and NIRVANA
\citep{ziegler&yorke97}, and a Godunov code, RAMSES
\citep{teyssier02}. The implementation of  shearing box simulations using
a Godunov code is described in detail in \citet{gardiner&stone05b} and was
subsequently applied using RAMSES \citep{fromang_ramses06}.

We  consider only the case $(c/U)^2=0.3$ here since this is the case we
will study in more detail in section~\ref{insta_num}. We perform
different 1D numerical simulations of wave propagation for different
wave amplitudes. In each case, the size of the box is set equal to the
wavelength of the wave. We used periodic
boundary conditions in shearing coordinates and followed the wave over many cycles.

 The point of these tests is to demonstrate that we can
successfully follow free propagating non linear waves. 
In particular we need to verify that this can be done under the
 conditions adopted in
section~\ref{insta_num} where we study the hydrodynamic stability
of the waves. For that reason, we present results obtained
with a single resolution having $320$ grid cells in the radial
direction. In these simulations, the timestep $\Delta t$ is defined by
\begin{equation}
\Delta t=C_0\frac{\Delta x}{\textrm{max}|v_x|+c}
\end{equation}
where $\Delta x$ stands for a grid cell size and $C_0$ is the Courant
number, which should be  positive and smaller than one for the CFL condition to be
satisfied \citep{numericalrecipes}. 

\subsubsection{Results}
We considered waves with two amplitudes. These are given by
$\rho_{max}/\rho_0=1.05$ and $\rho_{max}/\rho_0=1.8$. The results
obtained with ZEUS
and RAMSES are illustrated in figure~\ref{1d_tests}. Results obtained
using NIRVANA are very similar to the former. Each of the panels in
figure~\ref{1d_tests} shows the radial profile of the
density at different times measured in 
 wave periods. The upper panels illustrate results
obtained for the low amplitude case. The lower panels  illustrate results obtained
for the high amplitude case. In the former case, these are shown 
for $t=0,$ $t=250$ and $t=500.$ In the latter case, results 
are shown for  $t=0,$ $t=30$  and
$t=60.$ The left  panels
show results obtained with ZEUS with $C_0=0.5$, the central panels show
results obtained with ZEUS using $C_0=0.1$ and the right 
panels show results obtained with RAMSES with $C_0=0.7$.

\subsubsection{Effect of Courant number}
 Firstly, it is
apparent  that the ZEUS results obtained with $C_0=0.5$ are never acceptable
for long simulations. When the wave amplitude is small, we observed that
its profile is dispersed on timescales of hundreds of periods. When the
wave amplitude is large, small scale oscillations indicative of numerical
instability,  appear on timescales 
of tens of periods. 

As indicated in the central panels,
both problems are cured by decreasing the Courant
number. The wave drift relative to the centre of the box seen at large times in
both cases is due to the modification of the wave period as its
amplitude is slowly damped by numerical dissipation. 

Finally, as
shown in the right hand side panels, the results obtained with RAMSES
using $C_0=0.7$ are very stable and less dissipative than the ZEUS
results. In conclusion, despite differences in the codes, we found that
both Godunov and finite difference codes are able to follow the
evolution of these waves for a long enough timescale to perform the
simulations we will present in section~\ref{insta_num} provided the
Courant number is  chosen appropriately. For ZEUS or NIRVANA this means that
$C_0 \leq 0.1.$

\section{Stability of free waves}
\label{stability_section}
Having found exact nonlinear solutions corresponding to freely propagating 
density waves, an important issue is their stability. A possibility is that
such waves undergo parametric instability through interaction
with lower frequency inertial or gravity waves \citep{goodman93,pap05b}.
In addition to the production of these additional  waves, such
instabilities may result in the production of turbulence and the decay
of the initial wave amplitude and so affect its ability to propagate
over large distances.

Accordingly we now study the stability of  non linear density waves
to parametric decay into inertial waves (gravity waves being absent in
our model). This can be done while maintaining axisymmetry, ie. we
assume dependence of the state variables on $x$ and $z$ only.

\subsection{Stability analysis}

To study the stability of the nonlinear waves
described above we start from the equations of motion in Lagrangian form.
Following the discussion of section~\ref{analytic},  when the
vertical component of the gravitational acceleration, and thus 
vertical stratification  is neglected, these are seen to take the form
\begin{eqnarray}
\frac{D^2 X}{D t^2} + \Omega^2 (X-X_0) &=& -\frac{1}{\rho}
\frac{\partial P}{\partial x} \\
\frac{D^2 Z}{D t^2}  &=& -\frac{1}{\rho}
\frac{\partial P}{\partial z} \, .
\end{eqnarray}
We now want to study the evolution of a small perturbation of the
state defined by the unperturbed wave. To do so, we write
\begin{eqnarray}
X &=& X_0+\xi_x + \tilde{\xi}_x \\
Z &=& Z_0       + \tilde{\xi}_z \, ,
\end{eqnarray}
where $\bb{\tilde \xi} = (\tilde{\xi}_x , \tilde{\xi}_z)$ has a  small  magnitude in
comparison to that of 
$\xi_x \equiv \xi $ the displacement associated with the wave
used above. We therefore have $\Delta X=\tilde{\xi}_x$ and $\Delta
Z=\tilde{\xi}_z$, where $\Delta$ stands for the Lagrangian
perturbation. Upon noting that $\Delta$ and $D/Dt$ commute
\citep{lyndenbell&ostriker67}, we obtain
\begin{eqnarray}
\frac{D^2 \tilde{\xi}_x}{D t^2} + \Omega^2 \tilde{\xi}_x &=&
-\Delta\left(\frac{1}{\rho}\frac{\partial P}{\partial x}\right) =
-\Delta\left(\frac{\partial h}{\partial x}\right) \label{x_perturb} \\
\frac{D^2 \tilde{\xi}_z}{D t^2}  &=& -\Delta\left(\frac{1}{\rho}
\frac{\partial P}{\partial z}\right) = -\Delta\left(
\frac{\partial h}{\partial z}\right) \label{z_perturb} \, ,
\end{eqnarray}
where $h=c^2 \ln \rho$ is the enthalpy. Writing the Eulerian variation
of a quantity $f$ by $f'$, we have the relation $\Delta f= f' +
\bb{\tilde{\xi}}.\del f$ \citep{lyndenbell&ostriker67}, from which we obtain
\begin{eqnarray}
\Delta\left(\frac{\partial h}{\partial x}\right) &=& \frac{\partial
h'}{\partial x}+\tilde{\xi}_x \frac{\partial^2 h}{\partial x^2} \\
\Delta\left(\frac{\partial h}{\partial z}\right) &=& \frac{\partial
h'}{\partial z}
\end{eqnarray}
Using the continuity equation $\rho' + \del (\rho \bb{\tilde{\xi}})=0$,
we write
\begin{equation}
h'=c^2 \frac{\rho'}{\rho}=-\frac{c^2}{\rho}\frac{\partial \rho
  \tilde{\xi}_x}{\partial x}-\frac{c^2}{\rho}\frac{\partial \rho
  \tilde{\xi}_z}{\partial z} \, .
\label{continuity}
\end{equation}
Combining Eq.~(\ref{x_perturb}), Eq.~(\ref{z_perturb}) and
Eq.~(\ref{continuity}), we get
\begin{eqnarray}
\left(\frac{\partial}{\partial t}+v_x\frac{\partial}{\partial
  x}\right)^2 \tilde{\xi}_x + \Omega^2 \tilde{\xi}_x &=& c^2
\frac{\partial^2 \tilde{\xi}_x}{\partial x^2} + c^2 \frac{\partial^2
  \tilde{\xi}_z}{\partial x \partial z} + \frac{c^2}{\rho}
\frac{\partial \rho}{\partial x} \frac{\partial \tilde{\xi}_x}{\partial x}
\label{x_motion}\\
\left(\frac{\partial}{\partial t}+v_x\frac{\partial}{\partial
  x}\right)^2\tilde{\xi}_z \hspace{1.2cm} &=& c^2
\frac{\partial^2 \tilde{\xi}_z}{\partial z^2} + c^2 \frac{\partial^2
  \tilde{\xi}_x}{\partial x \partial z} + \frac{c^2}{\rho}
\frac{\partial \rho}{\partial x} \frac{\partial
  \tilde{\xi}_x}{\partial z} \label{z_motion}
\end{eqnarray}
where we used Eq.~(\ref{conv_der}) that defines the convective
derivative. Adopting a reference frame that moves with the speed $U$ of the
background wave (discussed in
  section~\ref{wave_equation}--\ref{ode_sol}), we also have
$\rho=\rho_0/(1+p)$ and $v_x=-U(1+p)$ where
$p=\partial \xi_x/\partial x$. These, along with the relation
  $v_y=-\frac{3}{2}\Omega X - \frac{1}{2}\Omega \xi$, fully determine
  the system. Finally, note that dissipation is neglected in our
  analysis. In Appendix~\ref{dissipation_sec}, we describe briefly how
  to incorporate it and the changes it produces. However, in the
  following, we assume the evolution is inviscid as it simplifies the
  analysis.

\subsection{The case $p=0$}
\label{case_p_zero}

In the absence of a background wave, $p=0$ and the previous system
reduces to the simpler form
\begin{eqnarray}
\left(\frac{\partial}{\partial t}-U\frac{\partial}{\partial
  x}\right)^2 \tilde{\xi}_x + \Omega^2 \tilde{\xi}_x &=& c^2
\frac{\partial^2 \tilde{\xi}_x}{\partial x^2} + c^2 \frac{\partial^2
  \tilde{\xi}_z}{\partial x \partial z} \label{radial_motion} \\
\left(\frac{\partial}{\partial t}-U\frac{\partial}{\partial
  x}\right)^2\tilde{\xi}_z \hspace{1.2cm} &=& c^2
\frac{\partial^2 \tilde{\xi}_z}{\partial z^2} + c^2 \frac{\partial^2
  \tilde{\xi}_x}{\partial x \partial z}  \label{vert_motion}
\end{eqnarray}
In this case, the radial displacement of a fluid element  can be taken to have the
form
\begin{equation}
 \bb{\tilde{\xi}} \equiv \bb{\tilde{\xi}}^n = ( \tilde
 {\xi}_x^{n},\tilde {\xi}_z^{n})  \propto e^{i(\omega_{I,n} t +n
 k_{x,w} x -k_z z)}
\label{mode_p0}
\end{equation}
where $n,$ as required by the radial periodicity condition, is an integer. 
\noindent Then Eq.~(\ref{radial_motion})
and Eq.~(\ref{vert_motion}) lead to the conditions 
\begin{eqnarray}
[(\omega_{I,n} - n\omega)^2-\Omega^2 -n^2k_{x,w}^2c^2]\tilde{\xi}_x^n +
n k_{x,w}k_z c^2\tilde{\xi}_z^n &=& 0 \label{inertial_x} \\
n k_{x,w} k_z c^2 \tilde{\xi}_x^n +[(\omega_{I,n}-n \omega)^2-k_z^2
  c^2]\tilde{\xi}_z^n &=& 0 \label{inertial_z}.
\end{eqnarray}
From these a dispersion relation can be derived in the form
\begin{equation}
[\omega_{I,n}-n \omega]^4-((n^2
k_{x,w}^2+k_z^2)c^2+\Omega^2)[\omega_{I,n}-n \omega]^2+\Omega^2 k_z^2
c^2=0 \, .
\label{disp_rel_iner}
\end{equation}
The low frequency branch of this dispersion relation corresponds to
inertial waves for which the frequency is given by 
\begin{equation}
(\omega_{I,n}-n \omega)^2=\omega_n^2=\frac{k^2c^2+\Omega^2}{2} \left(
    1 - \sqrt{1-\left[ \frac{4k_z^2\Omega^2c^2}{(k^2c^2+\Omega^2)^2}
    \right]} \right) \, ,
\label{inertial_freq_0}
\end{equation}
where $k^2=n^2 k_{x,w}^2+k_z^2$. In the large wavenumber limit, this
reduces to
$\omega_{I,n}-n \omega= \pm \omega_n = \pm \frac{k_z}{k} \Omega$ .
The frequency of these inertial waves is thus smaller than
$\Omega$ in the original non--comoving frame.

\subsection{The case $p \ne 0$}
\label{pne0}

When $p \ne 0$, but is in the small wave amplitude limit (i.e. $|p| \ll 1$),
 terms of higher than first order in $p$ may be neglected, so that
 the system of equations becomes
\begin{eqnarray}
\left[\frac{\partial}{\partial t}-U(1+p)\frac{\partial}{\partial
  x}\right]^2 \tilde{\xi}_x + \Omega^2 \tilde{\xi}_x &=& c^2
\frac{\partial^2 \tilde{\xi}_x}{\partial x^2} + c^2 \frac{\partial^2
  \tilde{\xi}_z}{\partial x \partial z} - c^2
\frac{\partial p}{\partial x} \frac{\partial \tilde{\xi}_x}{\partial x}
\label{base_x}\\
\left[\frac{\partial}{\partial t}-U(1+p)\frac{\partial}{\partial
  x}\right]^2\tilde{\xi}_z \hspace{1.2cm} &=& c^2
\frac{\partial^2 \tilde{\xi}_z}{\partial z^2} + c^2 \frac{\partial^2
  \tilde{\xi}_x}{\partial x \partial z} - c^2
\frac{\partial p}{\partial x} \frac{\partial
  \tilde{\xi}_x}{\partial z} \label{base_z}
\end{eqnarray}
In this limit, the background wave is linear and $p$ can be
written $p=\epsilon \cos (k_{x,w}x),$ with $\epsilon$ being a small parameter.
 If the frequency of the background wave,
$\omega$, lies between $\Omega$ and $2 \Omega$, then it is possible for
two inertial waves with wavenumbers $n k_{x,w}$ and $(n+1)k_{x,w}$ to interact with each other
by coupling through the original wave. Working in the frame comoving with the wave,
noting that the time dependence is separable,  we write the perturbed eigenvector 
as a linear combination of inertial waves in the
form
\begin{equation}
\bb{{\tilde \xi}} = (\bb{\tilde \xi}^n+\epsilon \bb{\tilde \eta}^n)
\,\, e^{i(\omega_{I,n}+\delta \sigma_n) t +
in k_{x,w}x - ik_z z} + (\bb{\tilde \xi}^{n+1}+\epsilon \bb{\tilde\eta}^{n+1})
\,\, e^{i(\omega_{I,n+1}+\delta \sigma_{n+1}) t + i(n+1) k_{x,w}x - ik_z z}
\label{MODE} \end{equation}
where the wave frequencies  are respectively
shifted by $\delta \sigma_n$ and $\delta \sigma_{n+1}$ which are both taken to be of order
$\epsilon.$ For the time dependence
 to be separable, so allowing the modes to interact, we require that 
\begin{equation}
\omega_{I,n}+\delta \sigma_n=\omega_{I,n+1}+\delta \sigma_{n+1} \, .
\end{equation}
Using Eq.~(\ref{inertial_freq_0}), if $\omega_{I,n}=\omega_n +n \omega$
and $\omega_{I,n+1}=-\omega_{n+1} +(n+1) \omega$, this is equivalent to
\begin{equation}
\omega_n+\omega_{n+1}=\omega+\delta \sigma_{n+1}-\delta
\sigma_n=\omega+\delta \omega \, .
\label{resonant_weak}
\end{equation}
Here $\omega_n, \omega_{n+1}, \omega $ and thus  $\delta \omega$ are given, while
(\ref{resonant_weak}) gives an interaction condition involving
 $\delta \sigma_{n}$ and $\delta \sigma_{n+1}.$
A further condition on the latter two, enabling the completion
of the solution,  is obtained as follows:

\noindent Four linear
equations relating the total of eight components 
of $\bb{\tilde \xi}^n , \bb{\tilde \xi}^{n+1}, \bb{\tilde \eta}^n $ and
$\bb{\tilde \eta}^{n+1}$ can  be obtained after substituting into  Eq.~(\ref{base_x}) and
Eq.~(\ref{base_z}) and  requiring that the  resulting   Fourier coefficients corresponding
to the terms in Eq.(\ref{MODE})  vanish. Setting $D_j = w_j^2-\Omega^2-j^2 k_{x,w}^2 c^2$
these are
\begin{eqnarray}
\epsilon \left[ D_n\tilde{\eta}_x^n +
 n k_{x,w} k_z c^2 \tilde{\eta}_z^n \right] + 2\omega_n \delta \sigma_n
 \tilde{\xi}_x^n + \epsilon (n+1) \left[\omega
 \omega_{n+1}-\frac{\omega^2}{2}-\frac{k_{x,w}^2c^2}{2}\right]
 \tilde{\xi}_x^{n+1} &=& 0 \label{first} \\
\epsilon \left[ D_{n+1}\tilde{\eta}_x^{n+1} +
 (n+1) k_{x,w} k_z c^2 \tilde{\eta}_z^{n+1} \right] - 2\omega_{n+1}
 \delta \sigma_{n+1}
 \tilde{\xi}_x^{n+1} - \epsilon n \left[\omega
 \omega_n-\frac{\omega^2}{2}-\frac{k_{x,w}^2c^2}{2}\right]
 \tilde{\xi}_x^n &=& 0 \label{second} \\
\epsilon \left[ (w_n^2-k_z^2 c^2)\tilde{\eta}_z^n +
 n k_{x,w} k_z c^2 \tilde{\eta}_x^n \right] + 2\omega_n \delta \sigma_n
 \tilde{\xi}_z^n + \epsilon (n+1) \left[\omega
 \omega_{n+1}-\frac{\omega^2}{2}\right] \tilde{\xi}_z^{n+1} -
 \frac{\epsilon}{2} k_z k_{x,w} c^2 \tilde{\xi}_x^{n+1} &=& 0 \label{third} \\
\epsilon \left[ (w_{n+1}^2-k_z^2 c^2)\tilde{\eta}_z^{n+1} +
 (n+1) k_{x,w} k_z c^2 \tilde{\eta}_x^{n+1} \right] - 2\omega_{n+1} \delta \sigma_{n+1}
 \tilde{\xi}_z^{n+1} - \epsilon n \left[\omega
 \omega_n-\frac{\omega^2}{2}\right] \tilde{\xi}_z^n +
 \frac{\epsilon}{2} k_z k_{x,w} c^2 \tilde{\xi}_x^n &=& 0\label{forth} 
\end{eqnarray}
Eq.~(\ref{inertial_x}) and Eq.~(\ref{inertial_z}) and their
counterparts for $n \rightarrow n+1$ give an additional  four  linear equations.
Being linear, in order that the total of eight equations be soluble,
their determinant must vanish. This leads to the requirement that 
\begin{equation}
\delta \sigma_n \delta \sigma_{n+1} = - \frac{\epsilon^2}{4} \frac{(A_n
    +B_n)(A_{n+1}+B_{n+1})}{C_n C_{n+1}}
    \frac{1}{\omega_n \omega_{n+1}} = -\sigma^2 \, ,
\label{growth_rate_zero}
\end{equation}
where
\begin{eqnarray}
\left\{ \begin{array}{ccc}
A_n &=& \left[(n+1)(k_z^2 c^2 -\omega_n^2)+n(\Omega^2+(n+1)^2 k_{x,w}^2
  c^2-\omega_{n+1}^2)\right]\omega
  \left(\omega_n-\frac{\omega}{2}\right) \\ 
A_{n+1} &=& \left[n(k_z^2 c^2 -\omega_{n+1}^2)+(n+1)(\Omega^2+n^2 k_{x,w}^2
  c^2-\omega_n^2)\right]\omega \left(\omega_n-\frac{\omega}{2}\right)
  \\ 
B_n &=& \frac{k_{x,w}^2 c^2}{2}(k_z^2 c^2 - (n+1) \omega_n^2) \\
B_{n+1} &=& \frac{k_{x,w}^2 c^2}{2}(k_z^2 c^2 + n \omega_{n+1}^2) \\
C_n &=& \Omega^2 + k_z^2 c^2 + n^2 k_{x,w}^2 c^2 - 2 \omega_n^2 \\
C_{n+1} &=& \Omega^2 + k_z^2 c^2 + (n+1)^2 k_{x,w}^2 c^2 - 2 \omega_{n+1}^2
\end{array} \right. \nonumber
\end{eqnarray}
Upon noting that $\delta \sigma_{n+1}=\delta \sigma_n+\delta \omega$, we find
\begin{equation}
\delta \sigma_n=\frac{\delta \omega}{2} \pm \sqrt{\frac{(\delta
    \omega)^2}{4}-\sigma^2}
\label{growth_rate}
\end{equation}
When $\delta \omega<2\sigma$, there is a non zero imaginary part to
both $\delta \sigma_n$ and $\delta \sigma_{n+1}$ and the amplitude of the
two coupled inertial modes grows: this is a parametric
instability. The maximum growth rate $\sigma$ is obtained when $\delta
\omega=0$, which occurs when the  resonant
condition  satisfied by the inertial waves is:
\begin{equation}
\omega_n+\omega_{n+1}=\omega \, .
\label{resonant_strong}
\end{equation}
When $\delta \omega$ does not vanish, the growth rate of the
instability is reduced, as also noted by  \citet{goodman93}.
We note  that on account of symmetry with respect to reflection
in the midplane, the growth rate does not change  when $k_z \rightarrow -k_z$. 
This enables solutions to be classified according to their symmetry 
with respect to reflection in the midplane.

\subsection{Unstable mode properties}
\label{mode_prop_sec}

\begin{figure}
\begin{center}
\includegraphics[scale=.9]{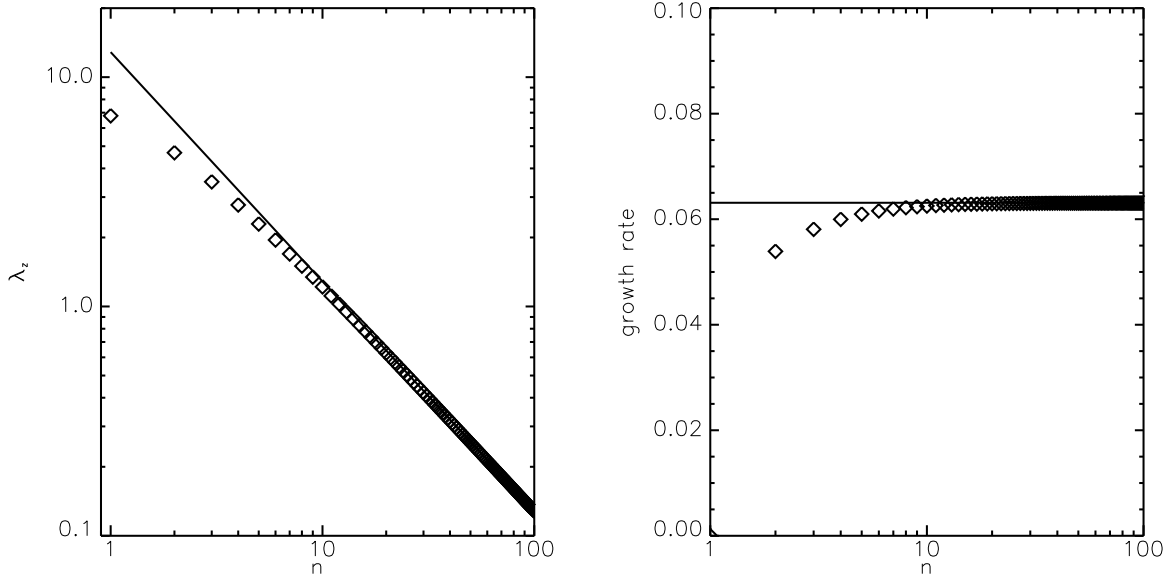}
\caption{Properties of the unstable modes of the parametric instability
for a wave characterized by $(c/U)^2=0.3$ and an amplitude such that
the maximum density $\rho_{max}/\rho_0 =1.05$. The two panels respectively show the
vertical wavelength, in units of the scale height,
 of the unstable mode and the growth rate in units of the wave period as a function of the
radial wavenumber $n$. In both panels, the solid lines represent the
large $n$ limit given by Eq.~(\ref{large_n_lim}) and
Eq.~(\ref{large_sigma_lim}).
Note that the mode vertical wavelengths become dense at large $n$
such that the requirement of being equal to the length
of a prescribed vertical domain such as the scale height when multiplied by some integer
can be approached arbitrarily closely.}
\label{mode_prop}
\end{center}
\end{figure}

In this section, we study the properties of modes that undergo
parametric instability for the special case $(c/U)^2=0.3$. We focus here
on modes that occur when   
the resonant condition given by
Eq.~(\ref{resonant_strong}) is satisfied exactly.

 For a given background wave amplitude,
the spatial period of the background wave is calculated as described
in section \ref{ode_sol}. 
The radial wavenumber of the background wave $k_{x,w}$ and the
wave frequency $\omega$ are then fixed.
 The instability couples two inertial waves
with radial wavenumbers $nk_{x,w}$ and $(n+1)k_{x,w}$. These  should
 satisfy the resonant condition, given by
Eq.~(\ref{resonant_strong}), and the dispersion relations, given by
Eq.~(\ref{disp_rel_iner}) and its counterpart with $n \rightarrow n+1.$
For  given values of $n,$ $\omega$ and $k_{x,w}$;  $\omega_n$,
$\omega_{n+1}$ and $k_z$ can be determined from these algebraic conditions.
At first we shall ignore any additional constraints on $k_z$ that may arise
from boundary conditions such as the requirement that the vertical wavelength should
be the scale height divided by an integer.

 We note that
the large wavenumber limit ($n \gg 1$) the solution is
given by
\begin{equation}
\omega_n=\omega_{n+1}=\frac{\omega}{2}  \, , \, \, 
k_z=n k_{x,w}\sqrt{\frac{\omega^2}{4\Omega^2 - \omega^2}} \, .
\label{large_n_lim}
\end{equation}
In the same limit, Eq.~(\ref{growth_rate})
for the growth rate gives
\begin{equation}
\sigma \sim \frac{\epsilon}{8} \Omega \left(\frac{c}{U}\right) \left(
\frac{\omega}{\Omega} \right)^2 (k_{x,w} H) \left(1+\frac{2}{(k_{x,w}
  H)^2}\frac{4\Omega^2 -\omega^2}{4\Omega^2}\right)
\, .
\label{large_sigma_lim}
\end{equation}
To illustrate the properties of the unstable modes,  in figure
\ref{mode_prop} we plot the growth rate
$\tilde{\sigma}=\sigma (2\pi/\omega)$ and vertical wavelength as a
function of $n$ for a wave
whose amplitude is such that $\rho_{max}/\rho_0=1.05$. 
It is seen that results obtained assuming the large $n$ limit applies
are always in good agreement with the actual results. 
We  comment that in the limit of large $n$ the modes become dense in the sense
that the requirement that the product of the vertical wave length
and some integer be the extent of a vertical domain or the
scale height can be satisfied with arbitrary accuracy.
 This follows from Eq. (\ref{large_n_lim}) and the fact that the rationals are dense.
 
\begin{figure}
\begin{center}
\includegraphics[scale=.9]{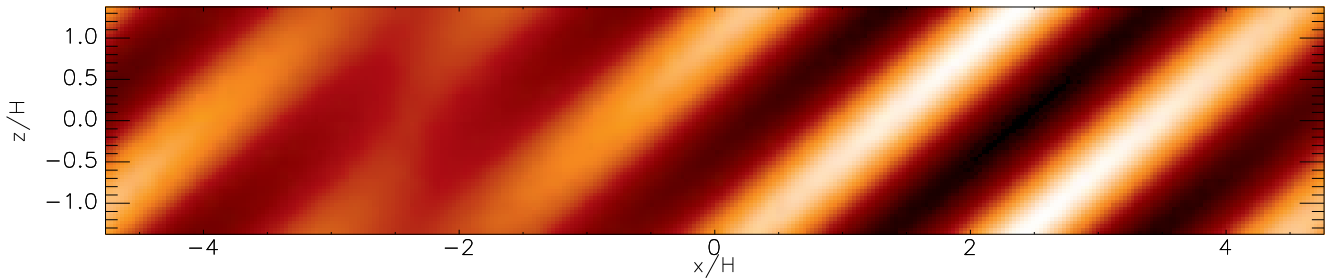}
\caption{Snapshot of the nondimensional vertical velocity in the $(r,z)$ plane
  showing the structure of the unstable mode of the parametric
  instability when $n=4$. The parameters of the background wave are
  $(c/U)^2=0.3$ and $\rho_{max}/\rho_0 = 1.05$.}
\label{growth_mode_th}
\end{center}
\end{figure}

Finally, the structure of the eigenmode is illustrated in a particular
case in figure~\ref{growth_mode_th}. This shows a snapshot of the
vertical velocity in the $(r,z)$ plane for a mode with 
$n=4$, in the case of a background wave for which
$(c/U)^2=0.3$ and $\rho_{max}/\rho_0=1.05$. The structure of this mode will
be compared in the following section with the results of numerical
simulations. 

\section{Numerical simulations}
\label{insta_num}

In this section, we present the results of numerical simulations
performed with ZEUS, NIRVANA and RAMSES. The goals are first to compare the
properties of the instability during the linear phase of the
parametric instability with the
analytical theory presented above. Then we investigate the stability
of strongly nonlinear waves. We also study the evolution of the
instability during its nonlinear phase. In  the next section we focus on
waves for which $(c/U)^2=0.3$.

\subsection{Small amplitude waves}
\label{unstrat_case}

\begin{figure}
\begin{center}
\includegraphics[scale=.9]{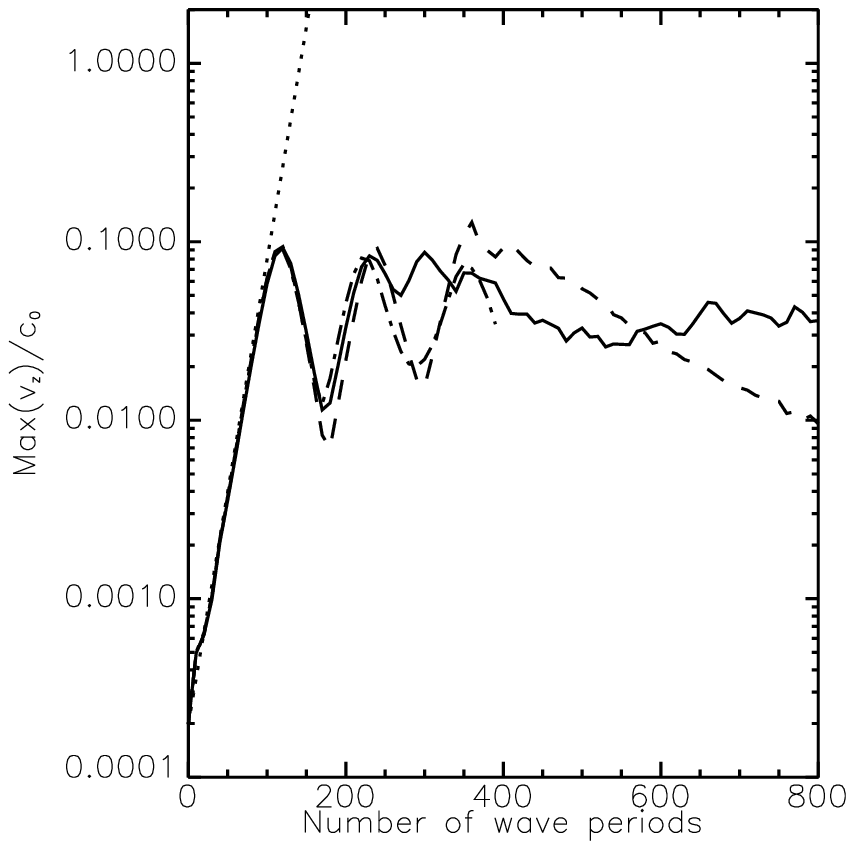}
\includegraphics[scale=.9]{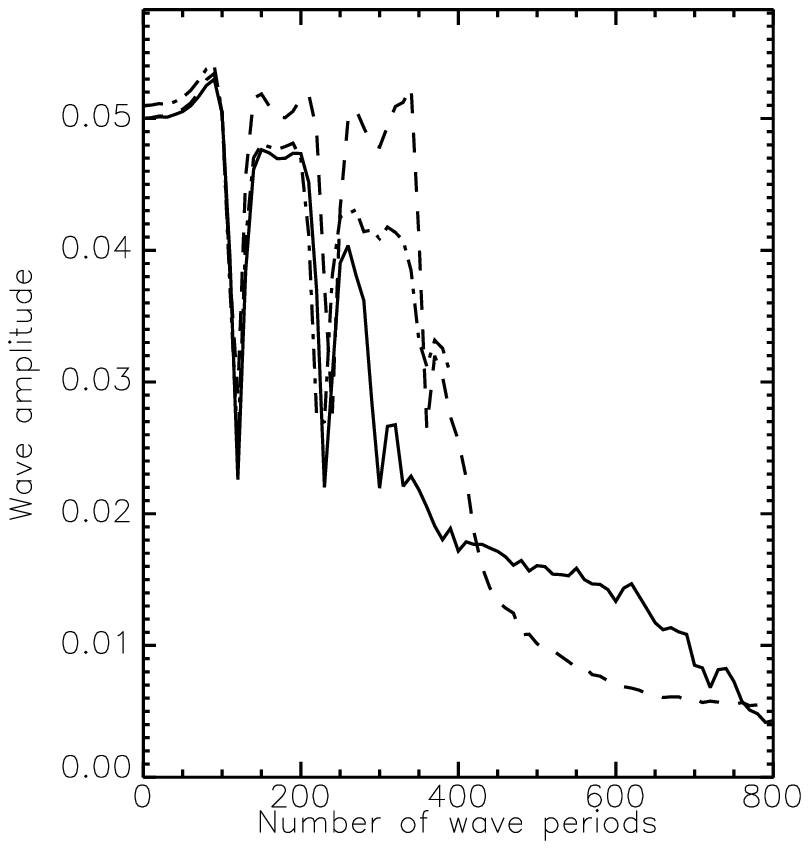}
\caption{The time history of the maximum value of the vertical velocity
  ({\it left panel}) and of the background wave amplitude $(\rho_{max}/\rho_0 -1)$
   ({\it right panel})
  as obtained with ZEUS ({\it solid line}), NIRVANA ({\it dot--dashed line})
  and RAMSES ({\it dashed line}). The dotted line shown in the left
   panel indicates the  behaviour expected from  linear theory.}
\label{time_hist_0.05}
\end{center}
\end{figure}

\begin{figure}
\begin{center}
\includegraphics[scale=.9]{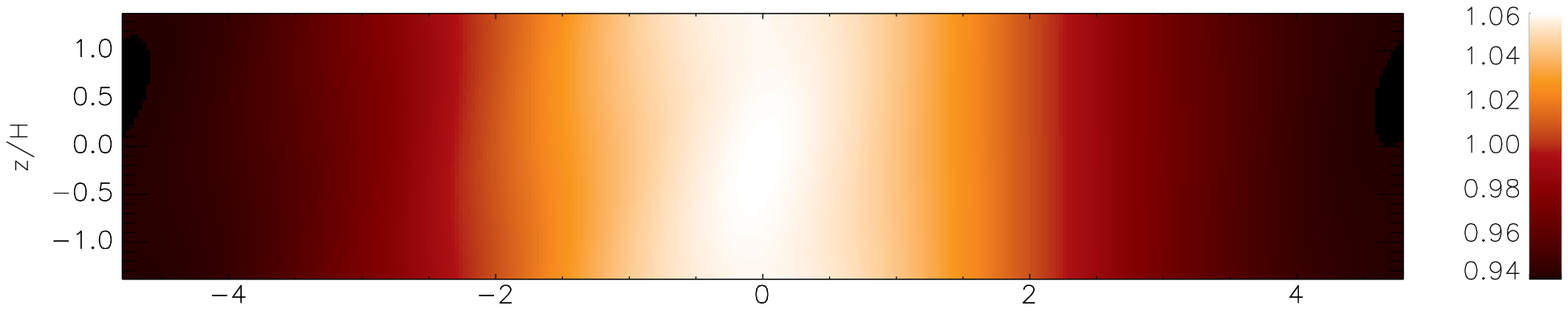}
\includegraphics[scale=.9]{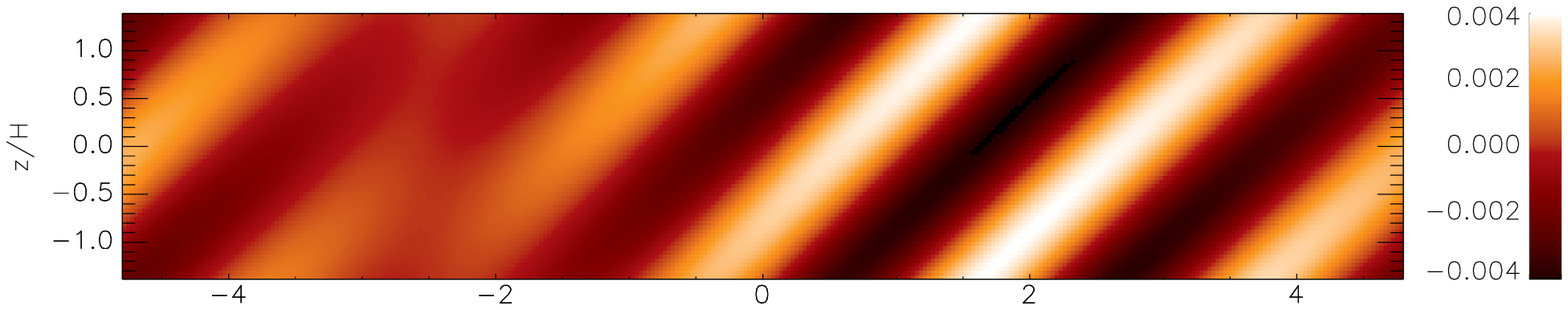}
\includegraphics[scale=.9]{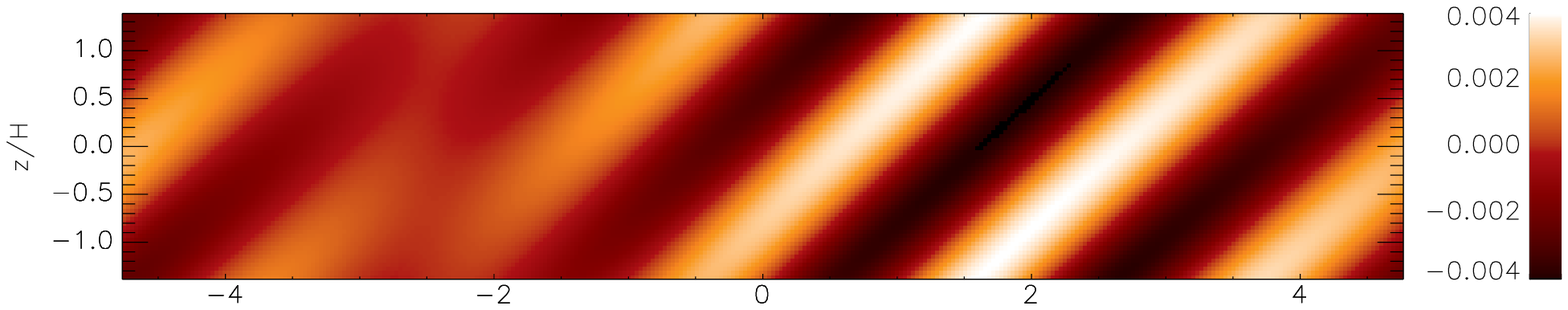}
\includegraphics[scale=.9]{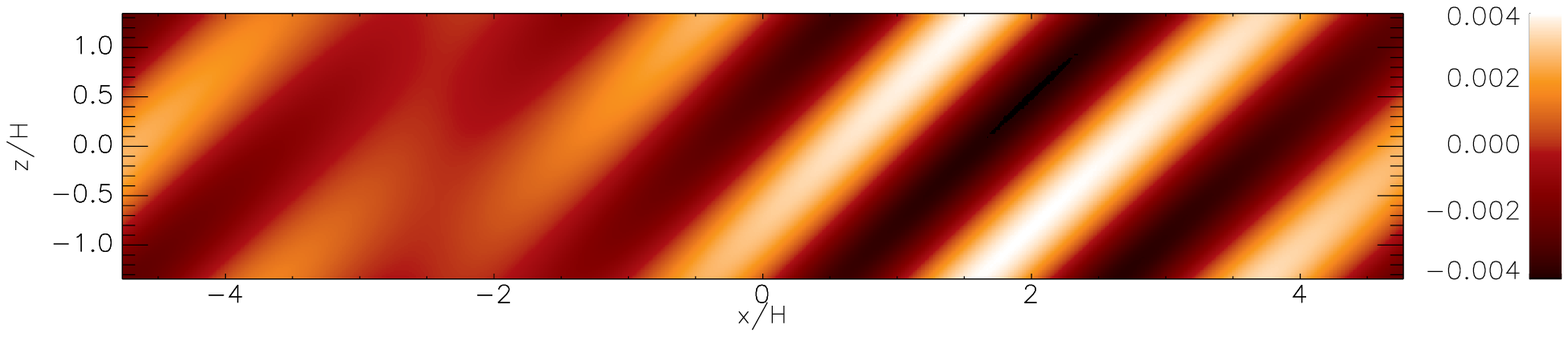}
\caption{Snapshots showing the structure of the parametric instability
  in the $(r,z)$ plane in the case of a linear background wave whose
  parameter are $(c/U)^2=0.3$ and $\rho_{max}/\rho_0=1.05$. The first two
  panels, both obtained with ZEUS, respectively show the density 
  (normalised by the mean density $\rho_0$) and
  vertical velocity distribution (normalised by the sound
  speed). The third and forth panels show
  snapshots of the vertical velocity obtained with NIRVANA and RAMSES
  respectively. Both are very similar to the ZEUS results. All the
  plots are obtained at $t=50$.}
\label{snapshots_0.05}
\end{center}
\end{figure}

We first study the case of a background wave with a small
amplitude such that $\rho_{max}/\rho_0=1.05$. The properties of the unstable
modes expected in that case were derived in section
\ref{mode_prop_sec}. From the solution of
Eq.~(\ref{wave equation}) we find that $T_w=9.6H$. The analytical
results (see  figure~\ref{mode_prop}) show that the most 
unstable modes have the largest wavenumbers. However, because of their
small scale structure, they will be difficult to represent
numerically. In order to illustrate the properties of
the parametric instability, we isolate the mode for which $n=4$.
This is achieved by adjusting the extent of the vertical domain appropriately.
 The results of
section \ref{mode_prop_sec} indicate that $\lambda_z=2.77H$ and 
$\tilde{\sigma}=0.06$ for that mode.

To follow its evolution, we performed 2D numerical
simulations in the $(x,z)$ plane with ZEUS, NIRVANA and
RAMSES. For the $n=4$ mode to develop, we chose a computational box
that covers the range $[-T_w/2,T_w/2]$ in $x$ and
$[-\lambda_z/2,\lambda_z/2]$ in $z$ (so that the radial extent
matches the wavelength of the background wave while the vertical
extent matches the vertical wavelength of the unstable mode). We used periodic
boundary conditions in both $x$ and $z$. Of course, other modes,
compatible with the boundary conditions 
(i.e. those which have a  ratio of box size to  vertical wavelength that
 is a large integer) could in principle grow as well. However, very
small scales modes will be strongly damped by numerical diffusion 
  (see appendix \ref{dissipation_sec}). In
practise, we found that they did not develop at the resolutions we
investigated. We also found that an accurate description of the $n=4$
mode requires roughly $64$ cells per wavelength. This translates into
a resolution $(N_x,N_z)=(320,64)$. For smaller resolutions, the growth
rate is reduced, although the evolution is qualitatively similar.

At time $t=0$, the density and velocity components are set equal to
the solutions of the system of differential equations given by
Eq.~(\ref{ode1}) and Eq.~(\ref{ode2}). A perturbation is also added to each
component of the velocity. If this perturbation was random, then
a combination of modes with vertical wavenumbers $+k_z$ and $-k_z$ would
grow with relative amplitudes depending on the  initial
conditions. This would make the comparison with the linear theory
less illustrative. Rather, we chose to add a perturbation that is
close to the mode with wavenumber $+k_z$ by adding a 
velocity perturbation of magnitude $\delta v$ such that
\begin{equation}
\delta v=\delta v_0 [-\sin(n k_{x,w}x-k_z z)+\sin((n+1)k_{x,w} x-k_z z)],
\end{equation}
where we took $\delta v_0=10^{-4} c$. In practise, we found this  procedure
 prevented the mode with wavenumber $-k_z$ from growing
during the course of our simulations.

The time history of the maximum value of the vertical velocity is
plotted in the left panel of figure \ref{time_hist_0.05}. The solid
line shows the results obtained with ZEUS, while the dot--dashed and
dashed line respectively correspond to the results obtained
with NIRVANA and RAMSES. All curves indicate an exponential growth
at early times. The evolution expected from  linear theory is also
plotted with the dotted line. There
is very good agreement between this   and the
numerical results, which themselves are in very good agreement. Results
from all three codes
show that the linear instability  saturates after about $100$ wave
periods, at which point the maximum vertical velocity has reached
about $10\%$ of the sound speed. This  is of the order of the radial 
velocity associated  with the background wave. This indicates that
  the linear instability saturates when the amplitude of the velocities of the
unstable inertial modes becomes large enough to perturb the structure of the
background wave. During the
nonlinear phase, the maximum value of $v_z$ at particular time is first found to
oscillate before it saturates (as obtained with ZEUS) or slowly decays
(as obtained with RAMSES). The oscillation is due to the linear
instability recurring quasi periodically between $t=100$ and $t=400$,
before the structure of the underlying wave is sufficiently
modified. Finally, we comment that the different evolution of the
simulations in the different codes is due to their different
dissipative properties. In order to illustrate the effect of the
instability on the wave structure, we plot the time history of its
amplitude on the right hand side panel of figure
\ref{time_hist_0.05} (with the same conventions as used on the
left hand side panel). Once again, there is a good agreement between the codes
during the linear phase of the evolution and during the recurring
instability phase. RAMSES displays a larger amplitude than ZEUS and
NIRVANA during that time. An additional run performed with NIRVANA
using twice the resolution gave similar results to those of RAMSES,
suggesting that this difference is due to the larger numerical
dissipation in the finite difference schemes. At late times, all codes
find that the wave amplitude is reduced to smaller and smaller
amplitudes, although there are differences due to the different small scale
dissipative properties. 

Figure \ref{snapshots_0.05} gives  plots of the structure of the
eigenmode obtained with the three different codes. The top
panel shows the density distribution, normalised by the mean
  density, in the $(r,z)$ plane after $50$
wave periods, as obtained with ZEUS. Being  in the linear phase,
it is hardly modified compared to the initial profile. The next three
panels, from top to bottom, represent the vertical velocity 
  (normalised by the sound speed $c$)
respectively with ZEUS, NIRVANA and RAMSES. They look almost identical
to each other and extremely similar to the expected theoretical
result shown in
figure \ref{growth_mode_th}.  

Taken together, all the results presented here demonstrate
a very good agreement between  the numerical and analytical
results. In the next section we investigate the properties of the
parametric instability that destabilizes waves of larger amplitude.

\subsection{Larger amplitude waves}
\label{large_amp_simus}

\begin{figure}
\begin{center}
\includegraphics[scale=.6]{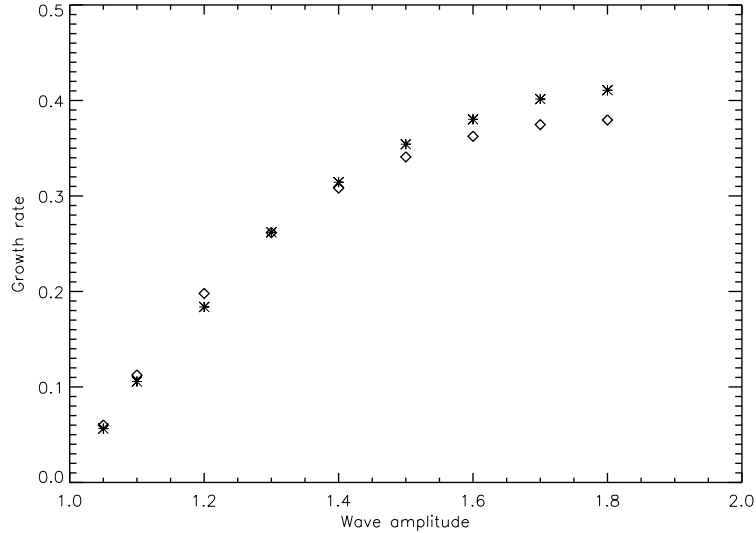}
\caption{Growth rate of the $n=4$ mode of the parametric instability
  for different background wave amplitudes that are characterized by
  $(c/U)^2=0.3$. The stars are determined by fitting the results of
  numerical simulations obtained with ZEUS. The circles are obtained
  using the results of the linear analysis (using
 Eq.~(\ref{growth_rate_zero}), the parameter $\epsilon$ is taken
  to be the amplitude of the Fourier mode whose wavelength equals that
  of the wave).}
\label{growth_rates}
\end{center}
\end{figure}

\begin{figure}
\begin{center}
\includegraphics[scale=.9]{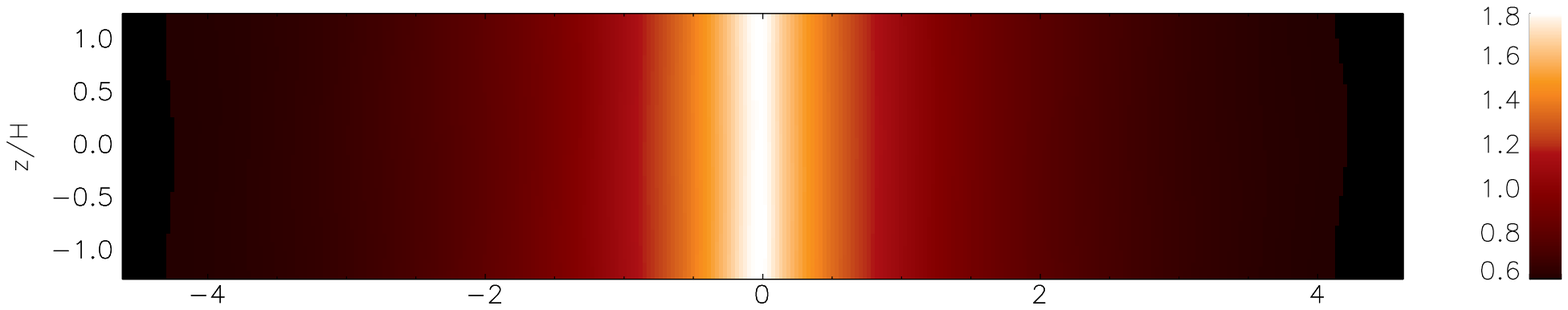}
\includegraphics[scale=.9]{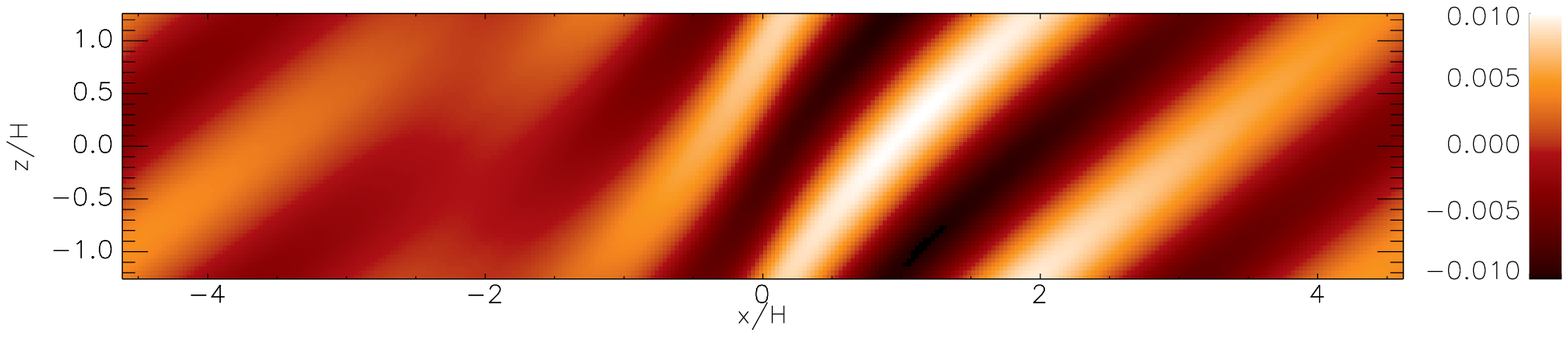}
\caption{Same as figure~\ref{snapshots_0.05}, but for a large
  amplitude background wave such that $\rho_{max}/\rho_0=1.8$. The snapshots
  are obtained at time $t=10$. Note the distorted structure of the
  eigenmode close to the density maximum. It results from the strong
  nonlinearities of the background wave at this location.}
\label{snapshots_0.8}
\end{center}
\end{figure}

Using ZEUS, we followed the evolution of waves of different amplitude.
We used the same set up (resolution, initial perturbation) as for the
small amplitude wave described above. In each case, the size of the
computational box is tuned to match the background wave period in the
radial direction and the wavelength of the $n=4$ mode in the vertical
direction. The results are illustrated in figure
\ref{growth_rates}. We show the growth rates $\tilde{\sigma}$ of the
instability measured during the linear
phase for wave amplitudes such that $\rho_{max}/\rho_0 = 1.05$, $1.1$,
$1.2$, $1.3$, $1.4$, $1.5$, $1.6$, $1.7$ and $1.8$. They can be compared
with the growth rates computed according to Eq.~(\ref{growth_rate_zero}).
 To calculate these,
the amplitude $\epsilon$ that appears in Eq.~(\ref{growth_rate_zero})
is taken to be equal to the amplitude of the Fourier mode 
with wavelength equal to the radial width of the box.
Figure \ref{growth_rates} shows that
there is  very good agreement between the results of the numerical
simulation and the analytical results for wave amplitudes smaller than
$\rho_{max}/\rho_0=1.5$. For larger amplitudes, the analytical estimates
tend to underestimate the growth rate slightly because of the strongly
nonlinear nature of the wave. This nonlinearity
modifies the structure of the eigenmode, as shown on figure
\ref{snapshots_0.8} for the case $\rho_{max}/\rho_0=1.8$. This figure is
similar to figure \ref{snapshots_0.05}: the top panel shows the
density distribution in the $(r,z)$ plane and the bottom panel
represents the vertical velocity. Both were obtained with ZEUS at time
$t=10$, well within the linear phase of the instability. A strong cusp,
already noted in figure \ref{wave_struct}, is seen almost
undisturbed on the upper snapshot. The lower one shows the
structure on the eigenmode. It is similar to the plots of figure
\ref{snapshots_0.05} although its pattern is distorted close to the
wave cusp. This is probably affecting the growth rate of the mode as
indicated in figure \ref{growth_rates}.

\begin{figure}
\begin{center}
\includegraphics[scale=.9]{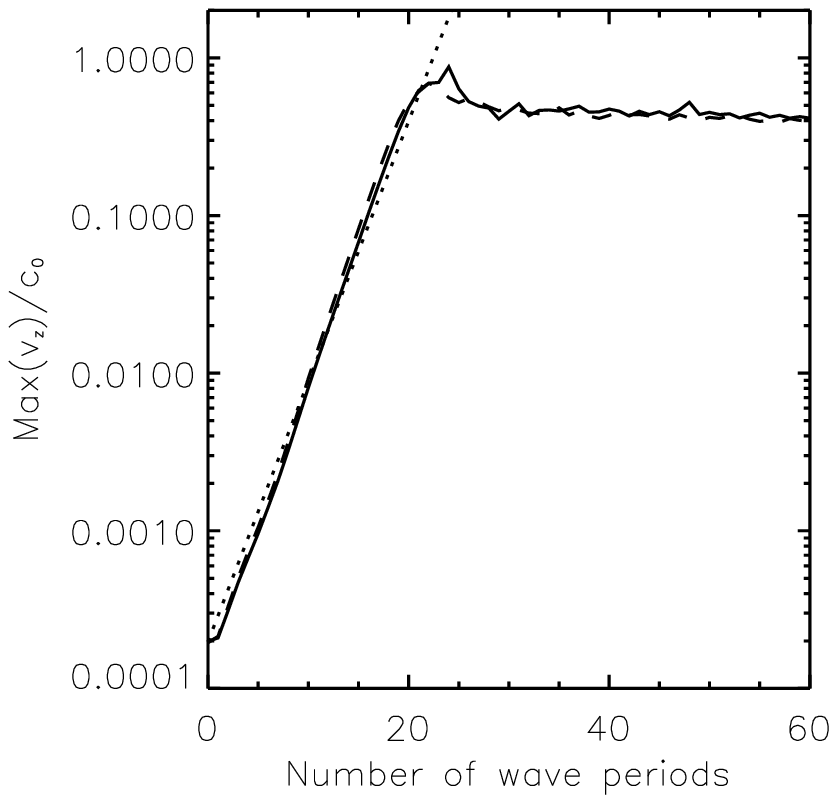}
\includegraphics[scale=.9]{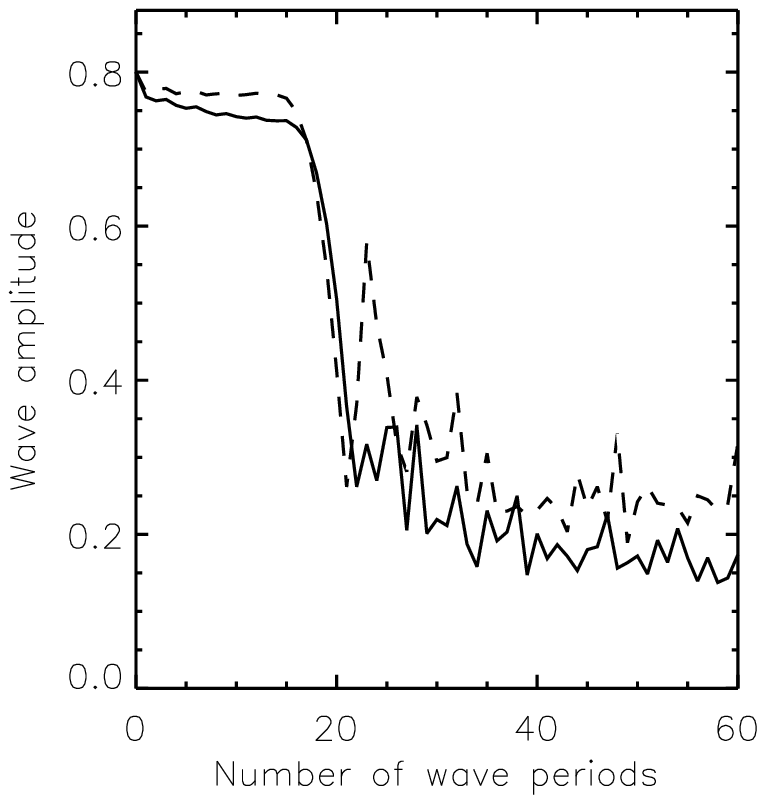}
\caption{Same as figure~\ref{time_hist_0.05}, but for a large
  amplitude background wave such that $\rho_{max}/\rho_0=1.8$ (only the
  ZEUS and RAMSES results are represented, respectively with the solid
  and dashed lines, while NIRVANA results are very similar). Note the
  faster growth of the instability compared to the smaller amplitude
  case.}
\label{time_hist_0.8}
\end{center}
\end{figure}

\begin{figure}
\begin{center}
\includegraphics[scale=.43]{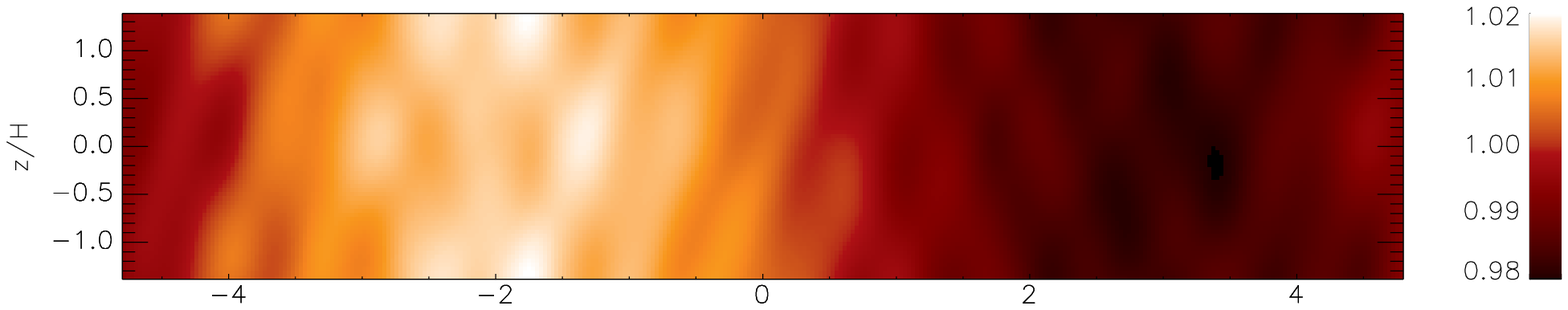}
\includegraphics[scale=.43]{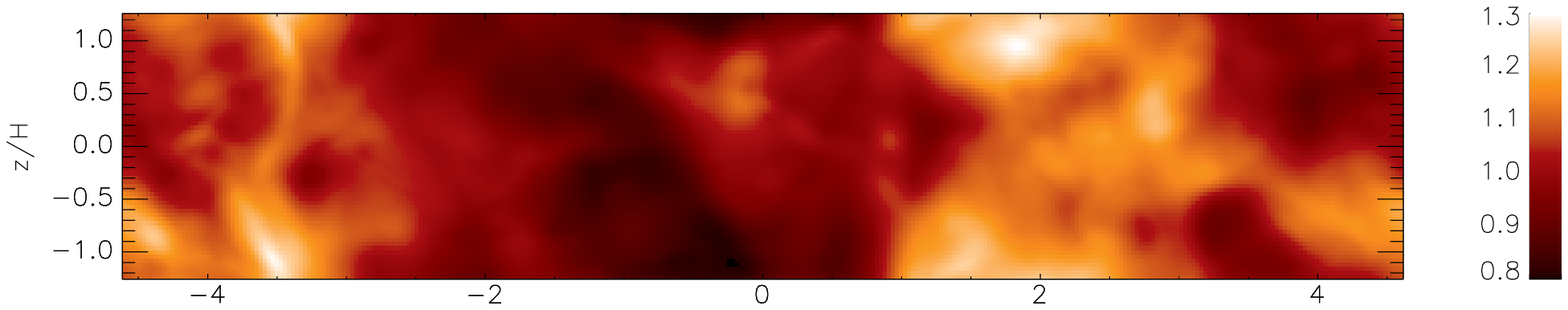}
\includegraphics[scale=.43]{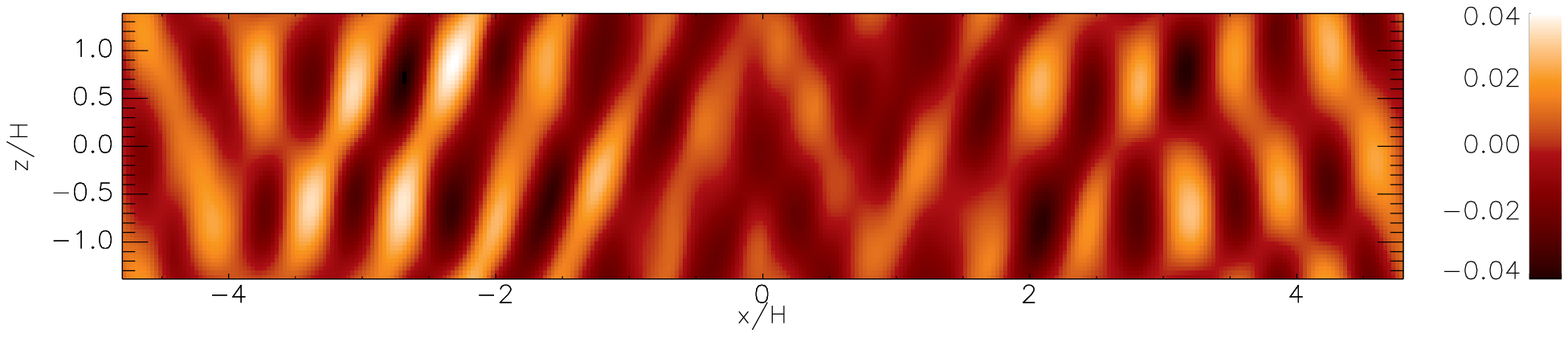}
\includegraphics[scale=.43]{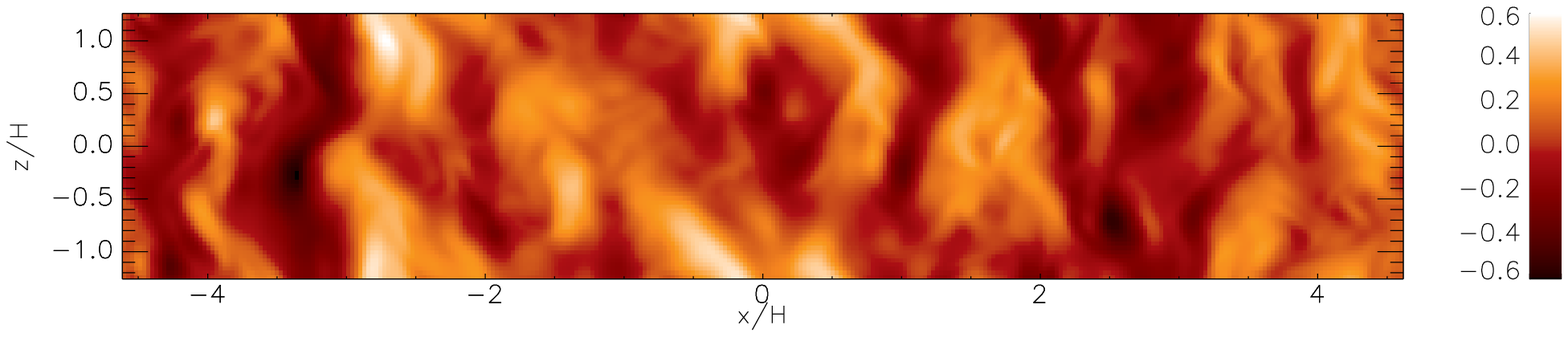}
\caption{Structure of the flow during the nonlinear stage of the
  parametric instability. Left panels correspond to a
  background wave amplitude $\rho_{max}/\rho_0=1.05$ and $(c/U)^2=0.3$. The
  upper figure plots $\rho/\rho_0$ in the $(r,z)$ plane, while the
  lower figure shows $v_z/c$. Both are
  plotted at $t=500$. The right hand side panels shows the same
  quantities at $t=27$ for a background wave amplitude
  $\rho_{max}/\rho_0$=1.8 and $(c/U)^2=0.3$.}
\label{struct_compar}
\end{center}
\end{figure}

The long term evolution of the instability in this strongly nonlinear
regime is further illustrated by figure \ref{time_hist_0.8}, which is
the 
equivalent of figure \ref{time_hist_0.05}. ZEUS results are plotted
using the solid line while RAMSES results are plotted with the dashed
line. Both are almost identical and results obtained using NIRVANA
displays a very similar 
behaviour. The evolution of the instability is qualitatively similar to
the small amplitude wave case, although the timescale for the
instability to saturate is much shorter, being $20$ wave periods when
$\rho_{max}/\rho_0=1.8$ as compared to  $100$ wave periods
when $\rho_{max}/\rho_0=1.05$ (once again, the linear instability
saturates when the perturbed velocities reach that of the background
wave, which happens to be close to the speed of sound in the large
amplitude case). In addition, no recurring instability phase is 
found before the amplitude of the wave starts to decrease. This is
better illustrated in figure \ref{struct_compar}. The left 
panels show the density structure ({\it upper panels}) and vertical
velocity ({\it lower panels}) obtained in the small amplitude case
($\rho_{max}/\rho_0$=1.05) at time $t=500$. Even at those late times,
although  smaller scale disturbances are involved, the
regular pattern characteristic of the parametric instability is still
visible. This is in contrast with the snapshots obtained in the large
amplitude case ($\rho_{max}/\rho_0$=1.8) at time $t=27$ and shown in the
right  panels of figure \ref{struct_compar}. The nature of
the flow is far more disorganized and looks more like turbulence in
that case. Finally, we want to comment that the system eventually
  goes back to equilibrium, with the wave amplitude being strongly
  damped. This might seem to disagree with figures
  \ref{time_hist_0.05} and \ref{time_hist_0.8}, where the vertical
  velocity saturates and shows very little decrease. This
  is because a vertical flow of alternating direction sets in at large
  times due to the periodic boundary conditions in the vertical
  direction and damps only very slowly. $v_z$ being only a function of
  $x$, this flow simply superposes to the equilibrium without 
  disturbing it. We emphasize here that this is only an artifact of the
  numerical setup.

\section{Stratified case}
\label{strat_section}

The results of the previous section give confidence that the nature of
the instability is properly captured by the analytical work presented
in section~\ref{stability_section}. However, this analysis neglected
the disk density vertical stratification, which is the focus of
the present section.

\subsection{Inertial wave properties}

The properties of inertial waves in stratified isothermal disks  have been
studied by \citet{lubow&pringle93}. We recall these below  for
the case $\gamma=1$ in their notation, as this is the case of interest
here.   
 Let $v^{'}_z(x,z)$ be the Eulerian perturbation of the
velocity in the vertical direction. Because of vertical
stratification, the $z$ dependence in $v^{'}_z(x,z)$ is no longer
harmonic as it was in  the unstratified case. Instead, it can be shown
to be of the form
\begin{equation}
v^{'}_z(x,y,z)=\tilde{v_z}(z) \exp [i(nk_{x,w}x-\omega_n t)] \, ,
\end{equation} 
where $v_z^{'}(z)$ is a polynomial in $z$ of the form:
\begin{equation}
\tilde{v_z}(z)=\sum_{m=0}^{n_z} c_m z^m \, .
\label{vzprime}
\end{equation}
 Here $n_z$ is given by
\begin{equation}
n_z+1=\frac{\omega_{n,n_z}^2}{\Omega^2} \left( 1- \frac{n^2
      k_{x,w}^2 c^2}{\omega_{n,n_z}^2-\Omega^2} \right) \, ,
\label{disp_rel}
\end{equation}
where the frequency of inertial waves, which constitute the low
frequency branch of this relation, is now denoted by  $\omega_{n,n_z}$
since it depends on the two integers $n$ and $n_z$. In addition, the
coefficients $c_m$ satisfy a recurrence relation, given by Eq.~(49) of
\citet{lubow&pringle93}. 
In the large $n_z$ limit, $\tilde{v_z}(z)$ oscillates with a period
\begin{equation}
\lambda_{z,n_z} \sim \frac{2\pi H}{\sqrt{n_z}}
\label{wavelength_strat}
\end{equation}
in the vertical direction in the vicinity of the equatorial plane.
Because of separability, two inertial waves with
the same $n_z$   but  different values of $n$ have the same vertical
structure. They can therefore interact together in the presence of the
background nonlinear sound wave when they satisfy the same resonant
condition as in the unstratified case:
\begin{equation}
\omega_{n,n_z}+\omega_{n+1,nz}=\omega \, .
\label{resonant_strat}
\end{equation}
As for the unstratified case, the spectral 
 frequencies $\omega_{n,n_z}$ are dense in the interval $[0,\Omega]$
(this can be demonstrated by showing that the difference
$\omega_{n+1,n_z}-\omega_{n,n_z}$ tends to zero as $n_z$ tends to
infinity): there will always be an infinite number of values of $n_z$ such
that Eq.~(\ref{resonant_strat}) is satisfied to a specified  accuracy.
But note that the modes are in fact {\it denser} in the stratified
 case than in the unstratified case.
This is because $n_z$ appears linearly in Eq. (\ref{resonant_strat}) whereas
the corresponding expression in the unstratified case would contain $n_z^2$
(see Eq.~(\ref{large_n_lim})). This has the consequence that the
 separation between neighbouring modes is less in the former case. A physical 
explanation for this is that this results from there being a
 continuous range of density scale heights in the stratified disk
 model compared to only one length scale in the unstratified
 case provided by the size of the computational box. This makes
 parametric instability more readily realised in the stratified case.

In a numerical simulation, however, only those modes for which the
wavelength is big enough (or $n$ small enough) to be well resolved will
have a non vanishing growth rate. This  limits the number of realisable modes
to those  that satisfy the resonant condition to  within a given
accuracy and are such that $n \leq n_{max}$. 

\subsection{Numerical simulations}

\begin{figure}
\begin{center}
\includegraphics[scale=0.9]{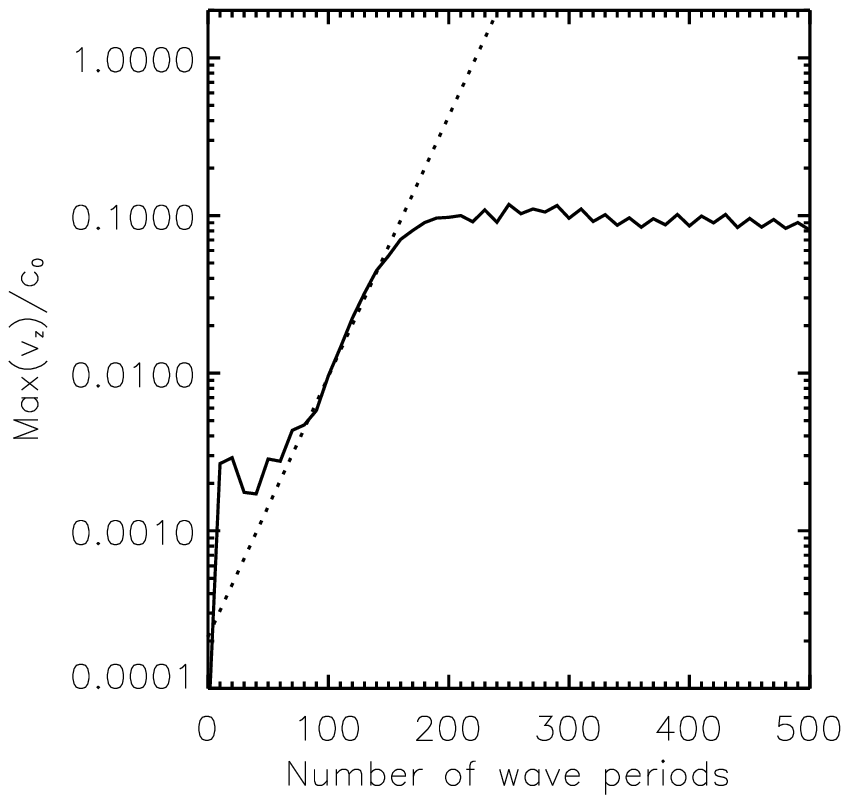}
\includegraphics[scale=0.9]{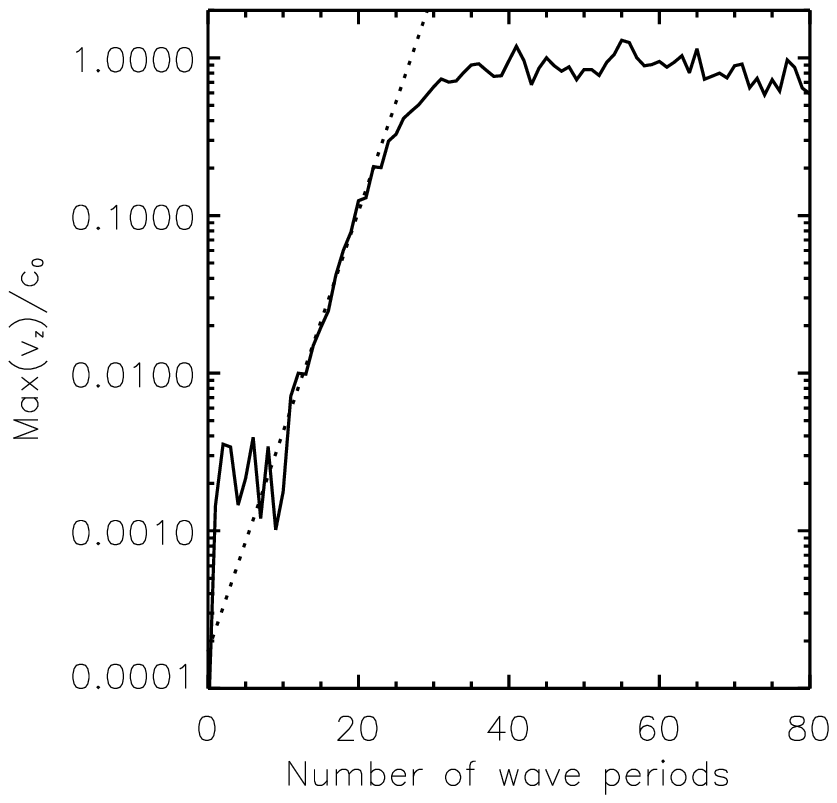}
\caption{Time history of the vertical velocity in the stratified
  model, for the small amplitude wave $\rho_{max}/\rho_0=1.05$ ({\it
  left panel}) and for the large amplitude wave
  $\rho_{max}/\rho_0=1.8$ ({\it right panel}). In both cases, the
  dotted line represents the exponential fit obtained during the
  linear phase of the instability. The growth rates that
  result are respectively $\tilde{\sigma}=0.038$ and $0.32$.}
\label{strat_history}
\end{center}
\end{figure}

\begin{figure}
\begin{center}
\includegraphics[scale=.43]{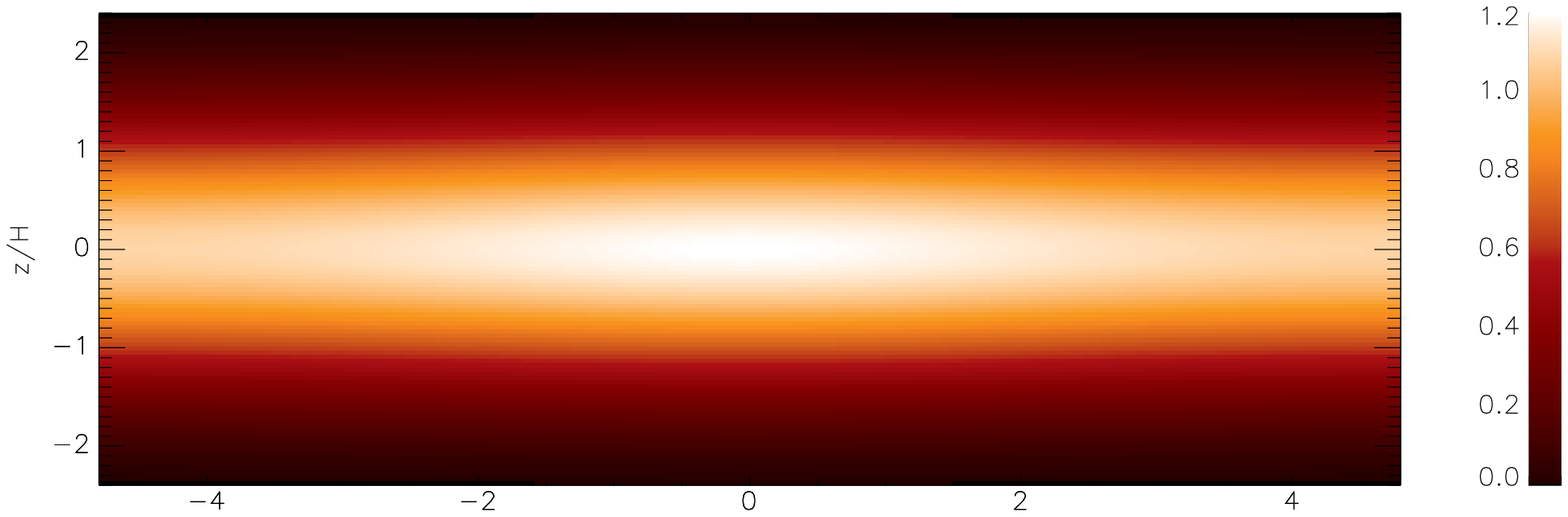}
\includegraphics[scale=.43]{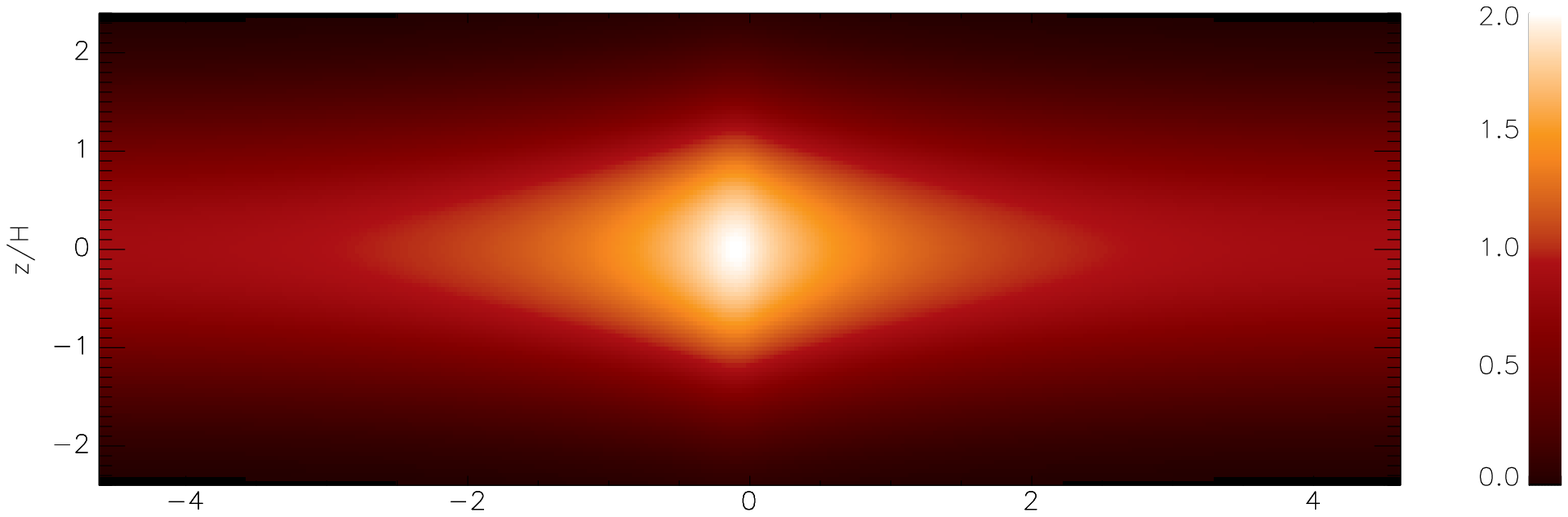}
\includegraphics[scale=.43]{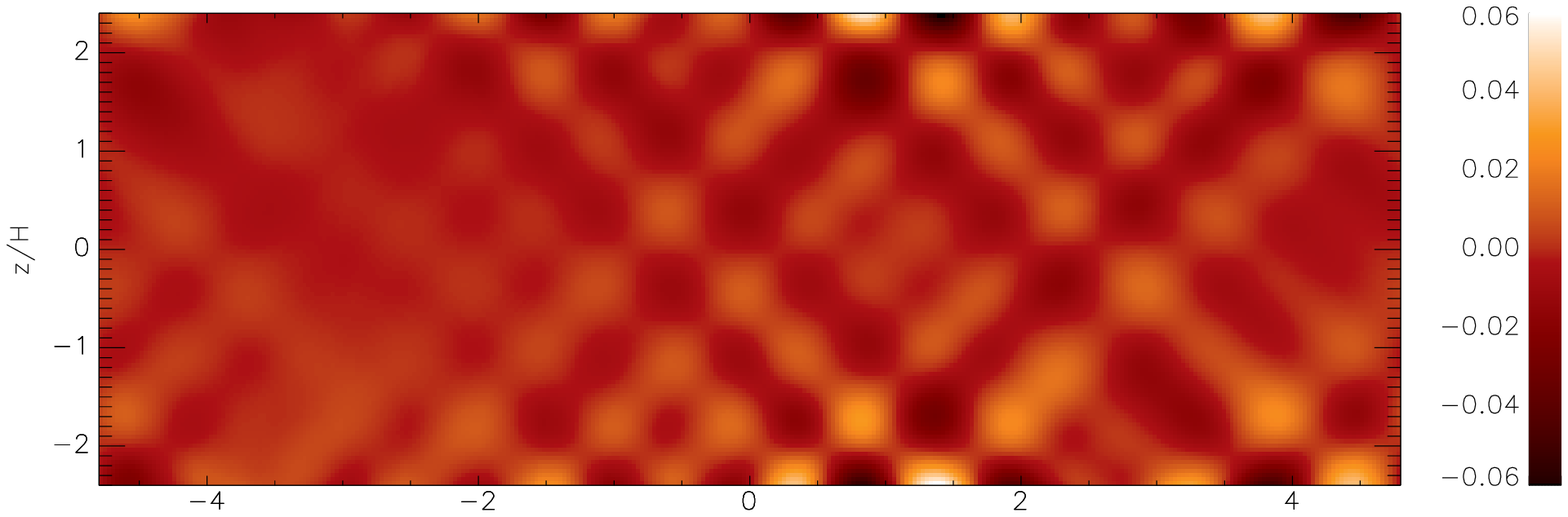}
\includegraphics[scale=.43]{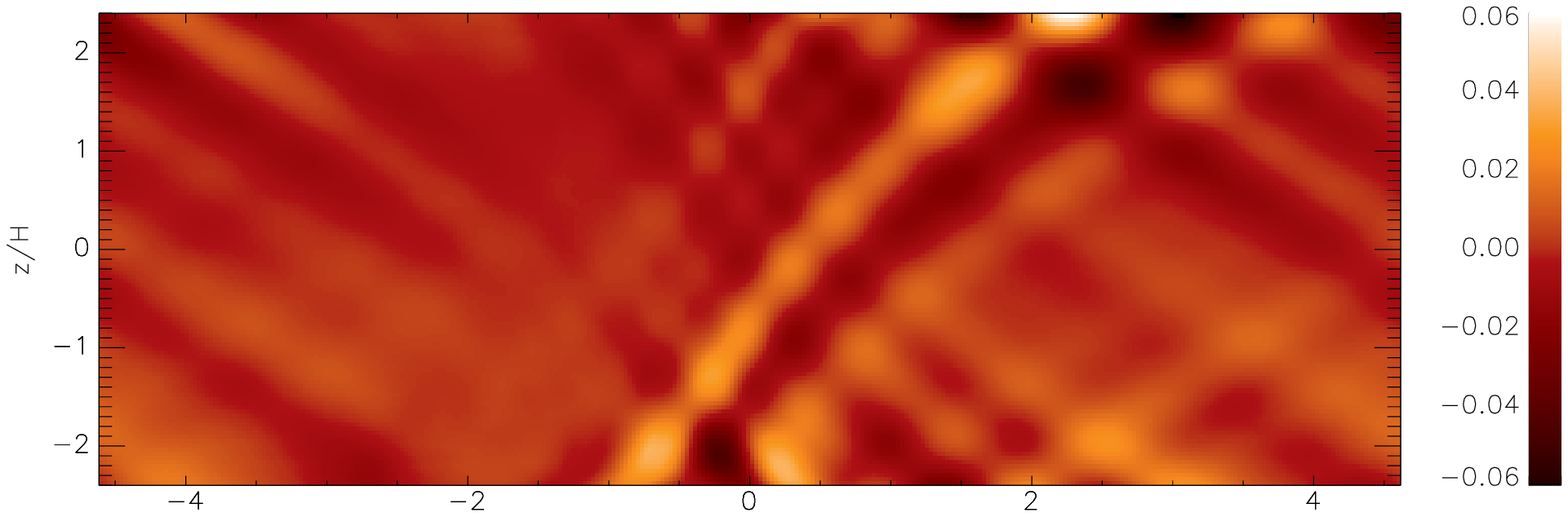}
\includegraphics[scale=.43]{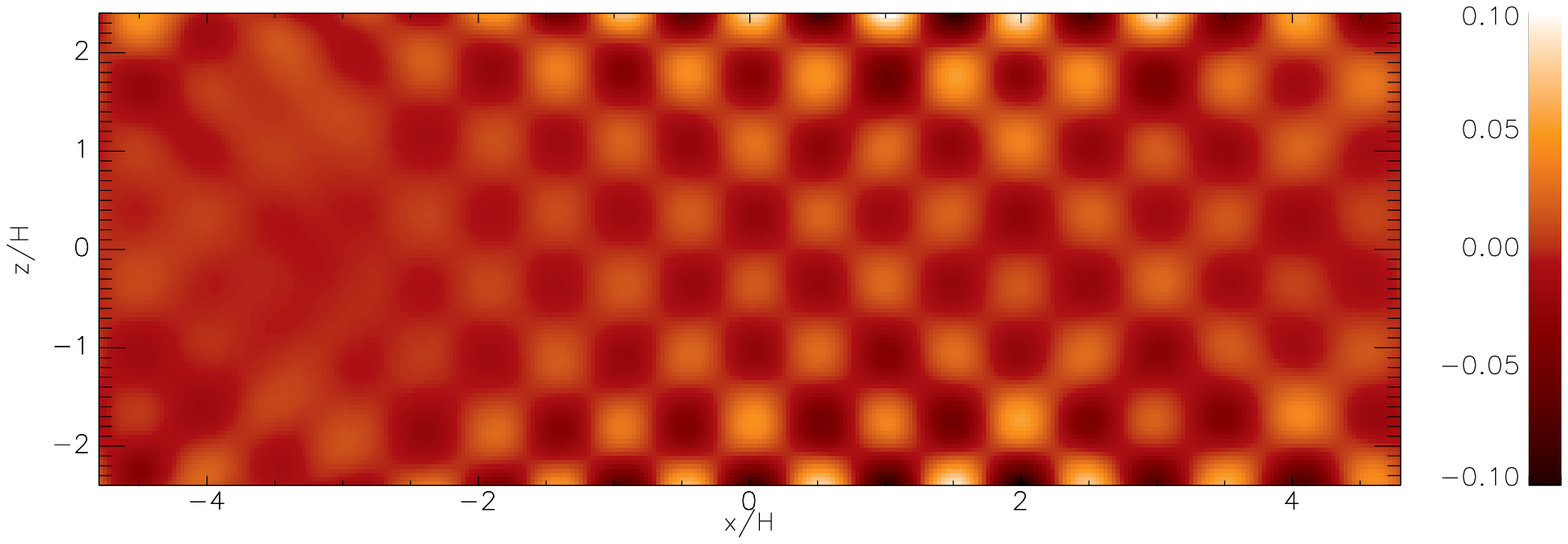}
\includegraphics[scale=.43]{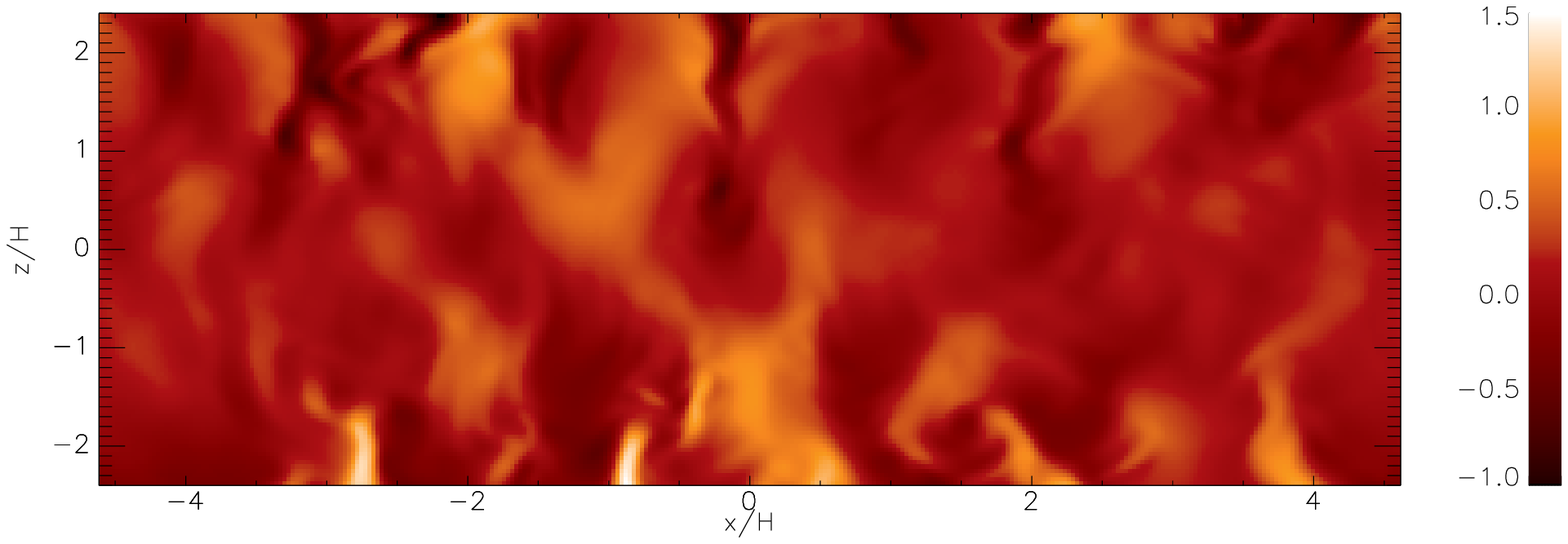}
\caption{Snapshots of the density and velocity in the $(x,z)$ plane
  showing the structure of the parametric instability in a stratified
  disk. The panels on the left correspond to the small amplitude wave,
  $\rho_{max}/\rho_0=1.05$, while the right hand side panels refer to the
  large amplitude wave $\rho_{max}/\rho_0=1.8$. On both sides, the
  top panels show the density (normalised by the mean density) during
  the linear phase (respectively at
  $t=150$ and $t=17$), the middle panels show the vertical velocity
  (normalised by the sound speed) at
  the same times and the bottom panels show the vertical velocity
  during the nonlinear phase (respectively at $t=450$ and $t=45$).}
\label{strat_snapshots}
\end{center}
\end{figure}

\begin{figure}
\begin{center}
\includegraphics[scale=.43]{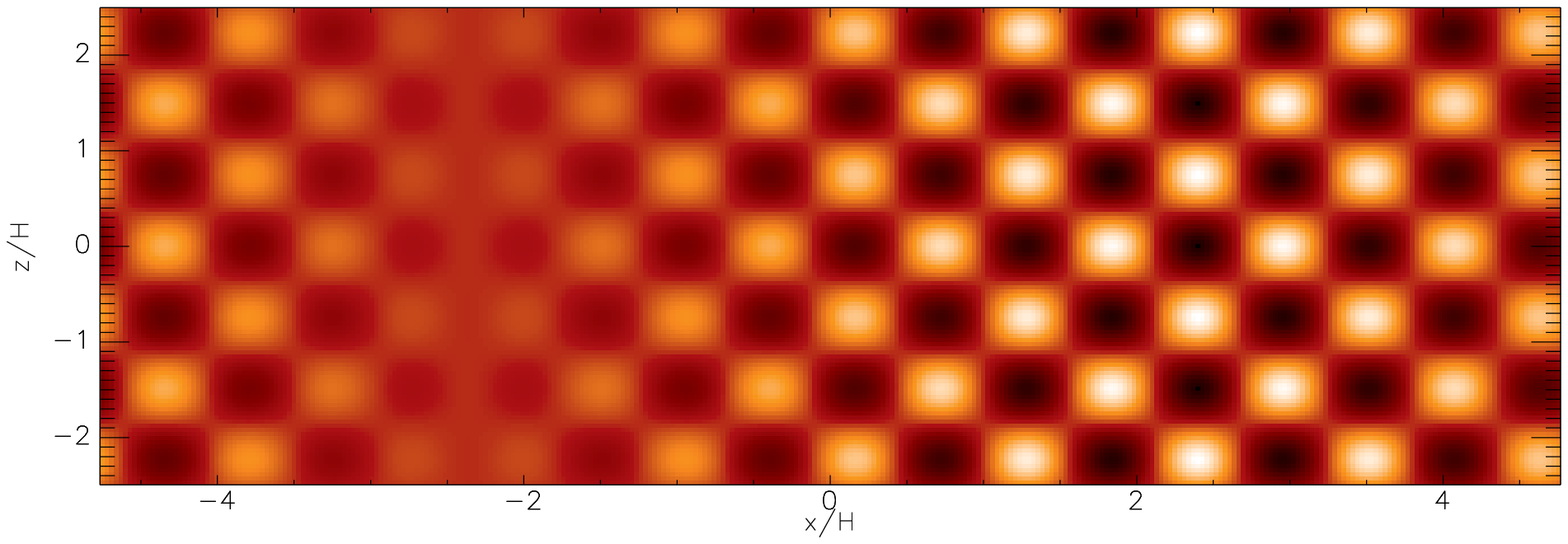}
\caption{Vertical velocity in the $(x,z)$ plane resulting from the
  $n=9$ mode only, calculated according to the linear results of
  section~\ref{stability_section} in the case of the small amplitude wave
  $\rho_{max}/\rho_0=1.05$. It is seen to be in good agreement with the
  numerical simulations (see the snapshots given by figure
  \ref{strat_snapshots}).}
\label{strat_theory}
\end{center}
\end{figure}

The simulations we describe in this section are the stratified
analogs of the simulations presented in section~\ref{insta_num} (we
only report results obtained with ZEUS here as we have shown good
agreement between the different numerical methods above). The
wave velocity is identical, $(c/U)^2=0.3$, and we consider only two 
wave amplitudes defined such that $\rho_{max}/\rho_0=1.05$ and
$\rho_{max}/\rho_0=1.8$. In our standard runs, the resolution is kept
the same in the $x$
direction, $N_x=320$, as is the extent of the computational domain, $x
\in [-T_w/2,T_w/2]$. In the vertical direction, the computational
domain is extended to cover $4$ scale heights on both sides of the
equatorial plane and the resolution is consequently increased to $193$
grid points. The initial density is set to be in hydrostatic
equilibrium in the vertical direction (this gives a Gaussian vertical
profile for the density) while the profile of the variables in the
radial direction is unchanged
compared to the unstratified case. Finally, we use periodic boundary
conditions in the
radial direction and outflow boundary conditions in the vertical
directions. At $t=0$, all components of the velocity are randomly
perturbed with an amplitude equal to $0.01 \%$ of the sound
speed. Note that this random initial condition does not break the symmetry
of the problem at $t=0$. Therefore, we expect modes with wavenumbers
$k_z$ and $-k_z$ to develop at the same time.

As in the unstratified simulations, the parametric instability is seen
to develop in both runs. This is illustrated in 
figure~\ref{strat_history} which shows the time history of the maximum value
of the vertical velocity in both cases (as before, time is measured in
units of the wave period). The left panel corresponds to
the small amplitude wave, $\rho_{max}/\rho_0=1.05$, and the right
panel to the large amplitude wave $\rho_{max}/\rho_0=1.8$. Similarly
to the unstratified case, the growth of the instability is faster in
the latter case. The growth rates, measured during the linear phase and
shown on figure \ref{strat_history} using a dotted line, are respectively
$\tilde{\sigma}=0.038$ and $\tilde{\sigma}=0.32$. Both are
comparable (although slightly smaller) than  the growth rates obtained in
section~\ref{large_amp_simus} and plotted in
figure~\ref{growth_rates}. We also checked that the amplitude of the
wave starts to decrease as the instability enters the nonlinear
regime and that the system goes back to hydrostatic equilibrium.
  The nonzero velocities seen at large times in
  figure~\ref{strat_history} are due to motions in the disk upper
  layers which are possibly long wavelengths acoustic modes excited
  by the parametric instability. Because of the low inertia of the
  gas, they take very long to damp. The structure
of the flow during the linear phase is illustrated in both
simulations by showing snapshots of different
quantities in the $(x,z)$ plane in figure~\ref{strat_snapshots}. The left
hand side snapshots correspond to the small amplitude wave case and
the right hand side snapshots to the large amplitude wave case (note
that both plots only cover the domain $[-2.5H,2.5H]$ in the
vertical direction, ie only a fraction of the actual computational box). On
both sides, the upper panel shows the distribution of the density
during the linear phase, respectively at times $t=150$ and $t=17$. As
in figure \ref{time_hist_0.8}, the cusp in the density is apparent in
the latter case. The middle panels in figure~\ref{strat_snapshots}  
represent the vertical velocity at the same times as the upper
panels. The stripped structure of the flow characteristic of the
parametric instability is obvious in both simulations. In the small
amplitude wave case, the structure of the unstable eigenmode is regular
while it is distorted in the neighbourhood of the cusp in the large
amplitude wave case. This is again similar to the results
obtained in the unstratified simulations. Also similar is the outcome
of the instability during the nonlinear regime. This is
illustrated by the lower panels of figure~\ref{strat_snapshots} which
show the structure of the vertical velocity in both cases,
respectively at times $t=450$ and $t=45$ (i.e. well into the nonlinear
regime). The small amplitude wave
simulation still displays a very regular structure while the entire
flow broke down into turbulence in the large amplitude wave
case. In conclusion, all of the results obtained in these stratified
simulations are very similar to the unstratified results of
section~\ref{insta_num}. 

\begin{table*}[t]\begin{center}\begin{tabular}{@{}cccc}\hline\hline
Mode number & $n$ & $n_z$ & precision \\
\hline\hline
1 & 9 & 21 & $2.3 \times 10^{-3}$ \\
2 & 10 & 26 & $4.6 \times 10^{-3}$  \\
3 & 7 & 13 & $6.3 \times 10^{-3}$  \\
4 & 10 & 25 & $7.6 \times 10^{-3}$  \\
5 & 8 & 17 & $8.2 \times 10^{-3}$  \\
6 & 4 & 4 &  $8.9 \times 10^{-3}$ \\
\hline\hline
\end{tabular}
\caption{Modes ranked according to the precision $\theta$ (see its
  definition in the text) within which they
  satify the resonant condition given by
  Eq.~(\ref{resonant_strat}). For each mode, labelled according to
  the first column, $\theta$ is given in the last column, while the
  second 
  and third columns respectively give the values of $n$ and $n_z$. The
  parameters of the model are such that $(c/U)^2=0.3$ and 
  the background wave has an amplitude $\rho/\rho_{max}=1.05$.}
\label{modes_properties}
\end{center}
\end{table*}

\begin{figure}
\begin{center}
\includegraphics[scale=.43]{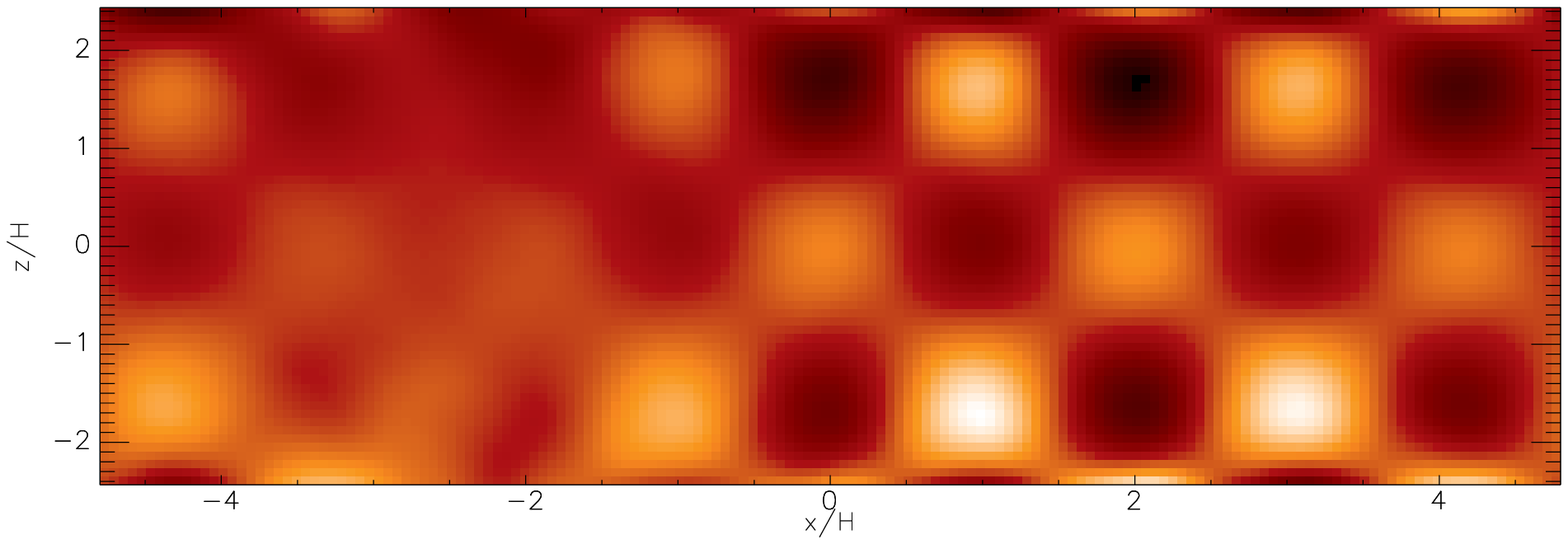}
\includegraphics[scale=.43]{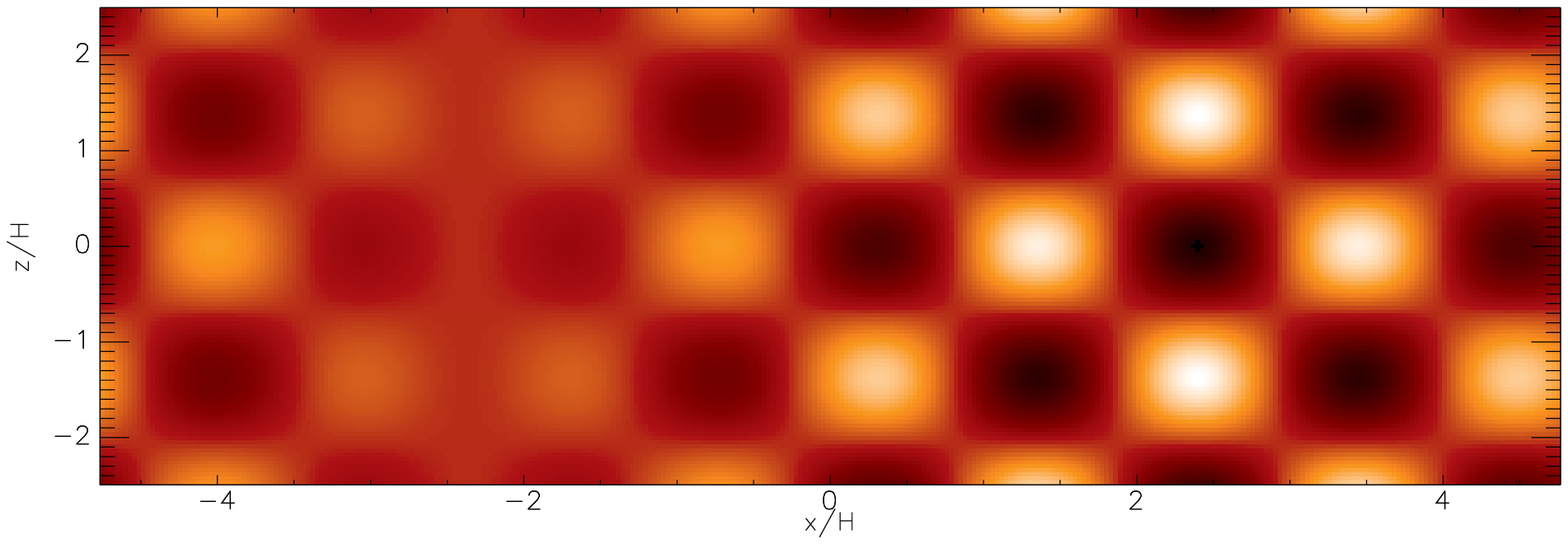}
\caption{Snapshot of the vertical velocity at time $t=200$ in the low
  resolution run: $(N_x,N_z)=(160,93)$ ({\it left panel}). The color
  scale spans the interval $[-0.15c,0.15c]$. Note the
  larger scale of the unstable eigenmode that grows in that case as
  compared to the standard run with twice that resolution. It is
  compared on the right panel with the structure of the $n=4$
  eigenmode as calculated according to the results of
  section~\ref{mode_prop_sec} (recall that both $n$ and $n+1$ contribute).
 Good agreement is found between the
  numerical and analytical results}
\label{strat_lowres}
\end{center}
\end{figure}

One may wonder, however, what determines the particular mode(s) that
grow in these stratified simulations. Unlike the unstratified case, no
vertical scale is determined by the size of the computational box. To
get some insight into this question, we now compare the results of the
small amplitude wave simulation to the linear analysis (performed in the {\it
  unstratified} box) presented in section~\ref{mode_prop_sec}. We expect the
latter to agree with the numerical results when large wavenumber modes are
excited. Indeed, the instability is truly local in that case and
should not be affected by
the vertical density stratification. In figure~\ref{strat_snapshots},
the dominant mode that emerges in the simulations seems to be such
that $n \sim 8$ to $10$. As an example, the structure of the flow
resulting from the mode $n=9$ is illustrated on
figure~\ref{strat_theory} (note that $k_z$ is determined
as the value that leads to the  satisfaction of 
 the  parametric resonance condition
 and both  modes with wavenumber $+k_z$ and
$-k_z$ are included).  It is roughly in agreement
with the left hand side snapshots of figure~\ref{strat_snapshots},
although some signs of the vertical stratification are seen in the
latter through an
increase of the velocity at large $z$. Figure~\ref{strat_snapshots} also
displays a signature of large scale variations in the radial direction,
possibly indicating that a smaller $n$ mode may be growing at the same
time (see below). The linear analysis indicates that the $n=9$ mode
should have a wavelength $\lambda_z=1.34H$, in rough agreement with
the simulation. However, it predicts a growth rate
$\sigma_{lin}=0.062$, larger than the numerically determined value
$\sigma_{num}=0.038$. This discrepancy is due
partly to numerical damping reducing the growth rate. It is also due
to the resonant condition being actually  only marginally satisfied, which
reduces the growth rate \citep{goodman93}. We quantify that accuracy
using the parameter $\theta$ which we define through
\begin{equation}
\theta=\frac{\omega_{n,n_z}+\omega_{n+1,nz}}{\omega}-1 \, .
\end{equation}
In the notation of section~\ref{pne0}, $\theta=\delta
\omega/\omega$. Eq.~(\ref{growth_rate}) states that the 
growth rate is unchanged when $\theta \ll \tilde{\sigma}/\pi$, while
the instability is suppressed when $\theta \ge
\tilde{\sigma}/\pi$. According to the linear results of
section~\ref{mode_prop_sec}, 
$\tilde{\sigma}/\pi \sim0.01$ for the small amplitude wave parameters
we are studying here. We therefore expect that only those
modes that have $\theta \le 0.01$ will grow with a significant growth
rate. Using Eq.~(\ref{disp_rel}), we searched for all the modes having
$n \le 10$ and $\theta \le 0.01$. The modes satisfying these
criteria are listed in Table~\ref{modes_properties}. The mode with the
smallest value of $\theta$ has $(n,n_z)=(9,21)$ and probably
corresponds to the mode described above and seen in the
simulations. In fact, Eq.~(\ref{wavelength_strat}) gives
$\lambda_{z,n_z}=1.37H$ for that mode, which is in good agreement both
with the results of the linear analysis mentioned above and to the
results of the numerical simulations. All the other modes, with $n \in [7,10]$ and increasing
values of $\theta$ probably contribute in a complicated way to the
growth of the parametric instability. Table 1 also reports the
existence of a mode
with $(n,n_z)=(4,4)$ and $\theta=8.9 \times 10^{-3}$ which probably
accounts for the large scale structure seen in
figure~\ref{strat_snapshots}. As the resolution increases, we expect
more modes with smaller wavelength to become important and to contribute to
the structure of the flow. This is also expected as a consequence
of numerical dissipation acting on the smallest scales (see appendix
\ref{dissipation_sec}). To illustrate these points, we performed a
simulation with the resolution decreased by a factor of two:
$(N_x,N_z)=(160,93)$. Everything else is kept identical. The structure
of the flow in that case is illustrated on the left panel of
figure~\ref{strat_lowres}. This is again a snapshot of the vertical
velocity in the $(x,z)$ plane at time $t=200$. In agreement with the
previous discussion, the growth of the parametric instability is
dominated by an eigenmode whose wavelength is larger than the high
resolution run. In fact, its structure probably results from the growth
of the $n=4$ mode listed in table~\ref{modes_properties}. To check
that hypothesis, the flow is compared in the right panel with the
structure of the $n=4$ mode  as computed according to the {\it unstratified}
linear analysis. The agreement between the analytical results and the
numerical simulations is very good.

\section{Conclusion}
\label{discussion_section}

In this paper, we have derived and solved
 a simple ODE which describes the structure
of nonlinear   density waves propagating in accretion disks. Such an
analysis is important for understanding the properties, stability and
dissipation of waves that propagate in disks in many situations: tidal
excitation by a companion, waves generated by the disk self--gravity
when it is massive enough or even waves generated by MHD turbulence
resulting from the MRI. For the analysis to remain tractable, we
adopted in the present paper the shearing box approximation and
restricted the equation of state to be isothermal and the wave to 
have no variation in the direction of the mean shear flow ie. to  be
axisymmetric. Because the waves were free and stationary
in a frame moving with the phase velocity, shocks
that would require a continual forcing or energy input 
 were not  considered here.

Using the solution of the ODE that describes the propagation of waves
of arbitrary amplitudes to provide initial conditions,
 we followed their  subsequent evolution 
in 1D using different numerical methods. A Godunov code, RAMSES,
was found to give very accurate results. Finite difference codes  such as 
ZEUS or NIRVANA were found to diverge from the correct solution or
develop small scale oscillations on long timescales. These problems
  were cured in our case  by decreasing the Courant   number to
$C=0.1$. However, we comment  that reducing the timestep only delays
the onset of the problems which were always found to develop provided
the waves   were evolved for long enough. 

We also studied the stability of the waves. When their frequency is
in the range $[\Omega,2\Omega]$, they can interact resonantly  with two linear
inertial waves  and  undergo a parametric
instability.  In this context we note that
this frequency range may be increased in stratified disks
if $\gamma \ne 1$ for perturbations allowing the existence of
$g$ modes \citep{lubow&pringle93}.
A wider class of waves would thus become candidates 
for parametric instability.  Neglecting vertical density stratification, we derived
an expression for the growth rate when the amplitude of the background
wave is small. This was found to be in good agreement with 2D
numerical simulations using RAMSES and ZEUS or NIRVANA (provided that
in the latter cases,  the
timestep was tuned as described above). Both unstratified and
stratified simulations were carried out. We found that the precise
structure of the mode that grows depends on the resolution we
used. This is because we used no physical dissipation and rather
relied on the resolution dependent
 numerical viscosity  to damp small
scale modes. Using these simulations, we
could study the nonlinear evolution of the instability.
For small amplitude waves the motion remained ordered but
large amplitude waves broke down to a turbulent state 
as might be expected from consideration of the related elliptical
instability in rotating flows.  In all cases,
we found that the amplitude of the wave was damped as the system
relaxed back to a stationary state. Of course, this is because no
external energy  that could keep the
instability going is provided.
 This differs from  realistic situations  that can occur, such as when the
disk is tidally forced by a companion.   Then the background
wave will be constantly forced. In such cases,  the external forcing 
will keep driving the instability which we expect to result in a
complicated quasi--steady state. We will study forced waves,
which also allow the incorporation of shocks, in a
future paper. 

\section*{ACKNOWLEDGMENTS}
SF acknowledges the important contribution of Romain Teyssier to the adaption
 of the numerical scheme of RAMSES to the shearing box equations.

\appendix
\section{The effect of a physical dissipation}
\label{dissipation_sec}

Here we consider the effect of a small constant kinematic viscosity $\nu$ on
the linear growth of the parametric instability. To do so, we study 
the properties of unperturbed inertial waves (see
section~\ref{case_p_zero}) when $\nu$ is nonzero. Note however that we
neglect the effect of the
dissipation on the background wave because its scale is much
larger. In the large wavenumber limit, we may neglect the effect of
possible stratification, the
inertial modes are nearly incompressible and Eq.~(\ref{radial_motion})
and (\ref{vert_motion}) can be written as
\begin{eqnarray}
\left(\frac{\partial}{\partial t}-U\frac{\partial}{\partial
  x}\right)^2 \tilde{\xi}_x + \Omega^2 \tilde{\xi}_x &=& c^2
\frac{\partial^2 \tilde{\xi}_x}{\partial x^2} + c^2 \frac{\partial^2
  \tilde{\xi}_z}{\partial x \partial z} + \nu \nabla^2 \left(\frac{\partial}{\partial t}-U\frac{\partial}{\partial
  x}\right) \tilde{\xi}_x \label{radial_motion_visc} \, , \\
\left(\frac{\partial}{\partial t}-U\frac{\partial}{\partial
  x}\right)^2\tilde{\xi}_z \hspace{1.2cm} &=& c^2
\frac{\partial^2 \tilde{\xi}_z}{\partial z^2} + c^2 \frac{\partial^2
  \tilde{\xi}_x}{\partial x \partial z}  + \nu \nabla^2 \left(\frac{\partial}{\partial t}-U\frac{\partial}{\partial
  x}\right) \tilde{\xi}_x  \label{vert_motion_visc} \, , 
\end{eqnarray}
when viscosity is taken into account. Using Eq.~(\ref{mode_p0}), these
two equations lead to:
\begin{eqnarray}
[\omega_n^2-n^2k_{x,w}^2c^2-i\nu k^2 \omega_n]\tilde{\xi}_x^n +
n k_{x,w}k_z c^2\tilde{\xi}_z^n &=& 0 \label{inertial_x_visc} \, , \\
n k_{x,w} k_z c^2 \tilde{\xi}_x^n +[\omega_n^2-k_z^2
  c^2-i\nu k^2 \omega_n]\tilde{\xi}_z^n &=& 0 \label{inertial_z_visc} \, ,
\end{eqnarray}
where $\omega_n=\omega_{I,n} - n\omega$ (note that we assumed that $k^2c^2
\gg \Omega^2$ in writing the above equations). Neglecting terms in order
$\nu^2$, these equations are in fact identical to their inviscid
counterparts, Eq.~(\ref{inertial_x}) and (\ref{inertial_z}), provided
$\omega_n$ is replaced by $\omega_{n,visc}=\omega_n-i\nu
k^2/2$. This indicates that inertial waves are damped on a timescale
\begin{equation}
\tau_{damp}=\frac{2}{\nu k^2} \, .
\label{damping_time}
\end{equation}
There will be a corresponding reduction in the growth rate of the
parametric instability. In numerical simulations performed with
Eulerian codes, we 
assume that the numerical dissipation has an effect similar to that
describe above, with a ``numerical'' viscosity $\nu_{num}$. For a
given resolution, $\nu_{num}$ is fixed and Eq.~(\ref{damping_time})
shows that modes with larger wavenumbers are damped more rapidly, in
agreement with our findings.

Another application of this result is to estimate a numerical Reynolds number
$Re$ for our simulations. $Re$ is given by
\begin{equation}
Re=\frac{cH}{\nu_{num}} \, .
\end{equation}
In section~\ref{unstrat_case}, we found that the linear growth rate of
the $n=4$ mode is unaffected by numerical dissipation, which
indicates that
\begin{equation}
\tau_{damp} \gg \frac{1}{\sigma} \, .
\label{compar}
\end{equation}
This gives an upper limit for the numerical viscosity of the codes and
a lower limit for the Reynolds number of our simulations. Introducing
the parameters of the simulation in Eq.~(\ref{compar}), we found $Re
\ge 10^3$. This is
consistent with the value reported by \citet{balbusetal96} for the
Reynolds number of Eulerian codes with similar resolutions.

\bibliographystyle{aa}
\bibliography{author}

\newcommand{\noopsort}[1]{}
\begin{thebibliography}{34}
\expandafter\ifx\csname natexlab\endcsname\relax\def\natexlab#1{#1}\fi

\bibitem[{{Arras} {et~al.}(2006){Arras}, {Blaes}, \& {Turner}}]{arrasetal06}
{Arras}, P., {Blaes}, O., \& {Turner}, N.~J. 2006, \apjl, 645, L65

\bibitem[{{Artymowicz} \& {Lubow}(1994)}]{artymowicz&lubow94}
{Artymowicz}, P. \& {Lubow}, S.~H. 1994, \apj, 421, 651

\bibitem[{Balbus \& Hawley(1998)}]{balbus&hawley98}
Balbus, S. \& Hawley, J. 1998, Rev.Mod.Phys., 70, 1

\bibitem[{{Balbus} {et~al.}(1996){Balbus}, {Hawley}, \& {Stone}}]{balbusetal96}
{Balbus}, S.~A., {Hawley}, J.~F., \& {Stone}, J.~M. 1996, \apj, 467, 76

\bibitem[{{Bate} {et~al.}(2002){Bate}, {Ogilvie}, {Lubow}, \&
  {Pringle}}]{bateetal02}
{Bate}, M.~R., {Ogilvie}, G.~I., {Lubow}, S.~H., \& {Pringle}, J.~E. 2002,
  \mnras, 332, 575

\bibitem[{{Fromang} {et~al.}(2006){Fromang}, {Hennebelle}, \&
  {Teyssier}}]{fromang_ramses06}
{Fromang}, S., {Hennebelle}, P., \& {Teyssier}, R. 2006, \aap, 457, 371

\bibitem[{{Gardiner} \& {Stone}(2005)}]{gardiner&stone05b}
{Gardiner}, T.~A. \& {Stone}, J.~M. 2005, in AIP Conf. Proc. 784: Magnetic
  Fields in the Universe: From Laboratory and Stars to Primordial Structures.,
  ed. E.~M. {de Gouveia dal Pino}, G.~{Lugones}, \& A.~{Lazarian}, 475--488

\bibitem[{{Goldreich} \& {Lynden-Bell}(1965)}]{goldreich&lyndenbell65}
{Goldreich}, P. \& {Lynden-Bell}, D. 1965, MNRAS, 130, 125

\bibitem[{{Goldreich} \& {Tremaine}(1979)}]{gt79}
{Goldreich}, P. \& {Tremaine}, S. 1979, ApJ, 233, 857

\bibitem[{{Goldreich} \& {Tremaine}(1980)}]{gt80}
{Goldreich}, P. \& {Tremaine}, S. 1980, ApJ, 241, 425

\bibitem[{{Goodman}(1993)}]{goodman93}
{Goodman}, J. 1993, ApJ, 406, 596

\bibitem[{Hawley \& Stone(1995)}]{hawley&stone95}
Hawley, J. \& Stone, J. 1995, Comput. Phys. Commun., 89, 127

\bibitem[{{Kato}(2001)}]{kato01}
{Kato}, S. 2001, PASPJ, 53, 1

\bibitem[{Kerswell(2002)}]{kerswell02}
Kerswell, R., R. 2002, Annual Review of Fluid Mechanics, 34, 83

\bibitem[{{Korycansky} \& {Pringle}(1995)}]{korycanskyetal95}
{Korycansky}, D.~G. \& {Pringle}, J.~E. 1995, \mnras, 272, 618

\bibitem[{{Larson}(1990)}]{larson90}
{Larson}, R.~B. 1990, \mnras, 243, 588

\bibitem[{{Lin} \& {Shu}(1964)}]{lin&shu64}
{Lin}, C.~C. \& {Shu}, F.~H. 1964, \apj, 140, 646

\bibitem[{{Lin} {et~al.}(1990{\natexlab{a}}){Lin}, {Papaloizou}, \&
  {Savonije}}]{linetal90a}
{Lin}, D.~N.~C., {Papaloizou}, J.~C.~B., \& {Savonije}, G.~J.
  1990{\natexlab{a}}, ApJ, 365, 748

\bibitem[{{Lin} {et~al.}(1990{\natexlab{b}}){Lin}, {Papaloizou}, \&
  {Savonije}}]{linetal90b}
{Lin}, D.~N.~C., {Papaloizou}, J.~C.~B., \& {Savonije}, G.~J.
  1990{\natexlab{b}}, ApJ, 364, 326

\bibitem[{{Lubow} \& {Ogilvie}(1998)}]{lubow&ogilvie98}
{Lubow}, S.~H. \& {Ogilvie}, G.~I. 1998, \apj, 504, 983

\bibitem[{{Lubow} \& {Pringle}(1993)}]{lubow&pringle93}
{Lubow}, S.~H. \& {Pringle}, J.~E. 1993, \apj, 409, 360

\bibitem[{{Lynden-Bell} \& {Ostriker}(1967)}]{lyndenbell&ostriker67}
{Lynden-Bell}, D. \& {Ostriker}, J.~P. 1967, \mnras, 136, 293

\bibitem[{Nelson {et~al.}(2000)Nelson, Papaloizou, Masset, \&
  Kley}]{nelsonetal00}
Nelson, R., Papaloizou, J., Masset, F., \& Kley, W. 2000, MNRAS, 318, 18

\bibitem[{{Ogilvie}(2001)}]{ogilvie01}
{Ogilvie}, G.~I. 2001, MNRAS, 325, 231

\bibitem[{{Ogilvie} \& {Lubow}(1999)}]{ogilvie&lubow99}
{Ogilvie}, G.~I. \& {Lubow}, S.~H. 1999, \apj, 515, 767

\bibitem[{{Papaloizou}(2005{\natexlab{a}})}]{pap05b}
{Papaloizou}, J.~C.~B. 2005{\natexlab{a}}, A\&A, 432, 757

\bibitem[{{Papaloizou}(2005{\natexlab{b}})}]{pap05a}
{Papaloizou}, J.~C.~B. 2005{\natexlab{b}}, A\&A, 432, 743

\bibitem[{{Papaloizou} {et~al.}(2004){Papaloizou}, {Nelson}, \&
  {Snellgrove}}]{papaloizouetal04}
{Papaloizou}, J.~C.~B., {Nelson}, R.~P., \& {Snellgrove}, M.~D. 2004, MNRAS,
  350, 829

\bibitem[{Press {et~al.}(1986)Press, Flannery, Teukolsky, \&
  Vetterling}]{numericalrecipes}
Press, W.~H., Flannery, B.~P., Teukolsky, S.~A., \& Vetterling, W.~T. 1986,
  Numerical Recipes: The Art of Scientific Computing (Cambridge Univ. Press)

\bibitem[{{Shu}(1992)}]{shu92}
{Shu}, F.~H. 1992, {Physics of Astrophysics, Vol. II} (Physics of Astrophysics,
  Vol.~II, by Frank H.~Shu.~Published by University Science Books, ISBN
  0-935702-65-2, 476pp, 1992.)

\bibitem[{{Teyssier}(2002)}]{teyssier02}
{Teyssier}, R. 2002, A\&A, 385, 337

\bibitem[{{Toomre}(1981)}]{toomre81}
{Toomre}, A. 1981, in Structure and Evolution of Normal Galaxies, ed. S.~M.
  {Fall} \& D.~{Lynden-Bell}, 111--136

\bibitem[{{Torkelsson} {et~al.}(2000){Torkelsson}, {Ogilvie}, {Brandenburg},
  {Pringle}, {Nordlund}, \& {Stein}}]{torkelssonetal00}
{Torkelsson}, U., {Ogilvie}, G.~I., {Brandenburg}, A., {et~al.} 2000, \mnras,
  318, 47

\bibitem[{{Ziegler} \& {Yorke}(1997)}]{ziegler&yorke97}
{Ziegler}, U. \& {Yorke}, H.~W. 1997, Computer Physics Communications, 101, 54

\end{thebibliography}

\end{document}